\documentclass[11pt,preprintnumbers,nofootinbib,prd]{revtex4}
\usepackage{graphicx}
\usepackage{amsmath}
\usepackage{epsfig}
\usepackage{mathrsfs}
\newcommand{\be}{\begin{equation}}
\newcommand{\ee}{\end{equation}}
\newcommand{\ba}{\begin{eqnarray}}
\newcommand{\ea}{\end{eqnarray}}
\newcommand{\notE}{E\kern-0.6em\hbox{/}\kern0.05em}
\newcommand{\notEt}{E_{T}\kern-1.21em\hbox{/}\kern0.45em}

\def\bi{\begin{itemize}}
\def\ei{\end{itemize}}

\begin{document}

\begin{flushright}
UCB-PTH-08/65\\
\end{flushright}

\title{Neutrino Masses, Baryon Asymmetry, Dark Matter and the Moduli Problem\\- {A Complete Framework}}
\author{Piyush Kumar}
\affiliation{Department of Physics, University of California,
Berkeley, CA 94720 USA\\and\\Theoretical Physics Group, Lawrence
Berkeley National Laboratory, Berkeley, CA 94720 USA}

%\date{\today}

\vspace{0.3cm}

\begin{abstract}

Recent developments in string theory have led to ``realistic"
string compactifications  which lead to moduli stabilization while
generating a hierarchy between the Electroweak and Planck scales
at the same time. However, this seems to suggest a rethink of our
standard notions of cosmological evolution after the end of
inflation and before the beginning of BBN. This epoch is crucial
for addressing the issues of neutrino masses, baryon asymmetry,
Dark Matter (DM) abundance and the moduli (gravitino) problem. We
argue that within classes of realistic string compactifications as
defined above, there generically exists a light modulus with a
mass comparable to that of the gravitino which is typically much smaller than the 
Hubble parameter during inflation. Therefore, it is destabilized and generates a large
late-time entropy when it decays. Thus, all known elegant
mechanisms of generating the baryon asymmetry of the Universe in
the literature have to take this fact into account.

In this work, we find that it is still possible to naturally
generate the observed baryon asymmetry of the Universe as well as
light left-handed neutrino masses from a period of Affleck-Dine
(AD) leptogenesis shortly after the end of inflation, in classes
of realistic string constructions with a minimal extension of the
MSSM below the unification scale (consisting only of right-handed
neutrinos) and satisfying certain microscopic criteria described
in the text. The AD mechanism has already been used to generate
the baryon asymmetry in the literature; however in this work we
have embedded the above mechanism within a framework well
motivated from string theory and have tried to describe the epoch
from the end of inflation to the beginning of BBN in a complete
and self-consistent manner. The consequences of our analysis are
as follows. The lightest left-handed neutrino is required to be
virtually massless. The moduli (gravitino) problem can be
naturally solved in this framework both within gravity and gauge
mediation. The observed upper bound on the relic abundance
constrains the moduli-matter and moduli-gravitino couplings since
the DM is produced non-thermally within this framework. Finally,
although not a definite prediction, the framework naturally allows
a light right-handed neutrino and sneutrinos around the
electroweak scale which could have important implications for the
nature of DM as well as the LHC.

\end{abstract}

\maketitle
\newpage
\tableofcontents

\section{Introduction}\label{intro}

Many theoretical and observational advances have been made to
uncover the mysteries of the very early Universe. Recent
cosmological observations from WMAP seem to favor an inflationary
phase of the Universe\footnote{In this work, we will assume that
the inflationary paradigm is correct.}. Although, there still does
not exist an agreed-upon microscopic theory of inflation, there
has been a lot of progress in this direction in recent years. The
theory of the primordial synthesis of nuclei - BBN, which starts
at a temperature of about an MeV, is also quite successful in
explaining observations. The evolution of the Universe after BBN -
leading to decoupling of matter and radiation, traditional matter
domination and large-scale structure formation, is also fairly
well understood.

However, not much is known about the epoch from the end of
inflation to the beginning of BBN. Theoretically, physics during
this epoch has been less studied as a whole (relatively speaking)
compared to the inflationary epoch. However, this epoch is quite
important for a number of reasons. For example, a crucial property
of the Universe, the existence of a baryon asymmetry, has to be
explained during this epoch, \emph{viz} after the end of inflation
and before BBN. This is because, inflation, in addition to
successfully diluting dangerous relics from the early past such as
monopoles, domain walls, cosmic strings, etc., also dilutes any
pre-existing baryon asymmetry. Therefore, the baryon asymmetry has
to be generated after the inflationary epoch. The existence of
baryon asymmetry requires that the Sakharov criteria be satisfied.
The three most popular ways of satisfying these by a) the
Affleck-Dine mechanism, b) out-of-equilibrium decay of a heavy
particle (as in GUT baryogenesis, thermal leptogenesis, resonant
soft leptogenesis) and c) during the electroweak phase transition
(as in electroweak baryogenesis) all happen before BBN. Various
particle-physics models of baryogenesis in this epoch
incorporating the above mechanisms, especially thermal
leptogenesis \cite{Fukugita:1986hr}, resonant soft leptogenesis
\cite{D'Ambrosio:2003wy}, affleck-dine baryogenesis
\cite{Affleck:1984fy} and leptogenesis \cite{Hamaguchi:2001gw} and
electroweak baryogenesis \cite{Cohen:1993nk} have been considered
in the literature. The origin of neutrino masses can be linked to
the generation of baryon asymmetry in models of leptogenesis,
providing an opportunity to solve both outstanding problems at the
same time. In this sense, the framework of leptogenesis is quite
appealing.

Many beyond-the-Standard-Model (BSM) particle physics models well
motivated from a microscopic theory such as string theory, also
have additional scalar particles known as ``moduli". These moduli
are scalar fields which couple very weakly to the visible sector
and scale like ordinary matter. Thus, they typically dominate the
energy density of the Universe and could decay after BBN spoiling
its successes, and at the same time greatly diluting any
previously generated baryon asymmetry.
Therefore, this provides a serious constraint to all existing mechanisms for producing the
baryon asymmetry. The ``cosmological moduli problem" is therefore
quite undesirable. In supersymmetric extensions of the
Standard-Model, the overproduction of gravitinos can cause similar
problems. In addition, the ``standard" picture in which the Universe is
radiation dominated during the whole epoch can be significantly
altered in the presence of moduli. In particular, Dark Matter (DM)
particles, instead of being produced during a phase of thermal
equilibrium, are typically dominantly produced non-thermally, via
the direct decay of moduli. However, this can lead to further
problems since it is easy to produce too much dark matter compared
with what we observe today. To summarize, therefore, understanding
the epoch starting from the end of inflation to the beginning of
BBN is extremely crucial to addressing all the above issues in a
systematic and holistic manner.

These problems have been known for a long time, various 
aspects of which have been discussed in the literature in the context of supergravity and string 
theory \cite{Banks:1993en}. However, 
early investigations of these issues, although important, were not very concrete 
as moduli stabilization in string compactifications and the resulting spectra of moduli was not well understood. With great improvement in our understanding of moduli stabilization in recent years, these issues warrant a careful investigation in light of the new developments.   

The aim of this work is to do precisely that - address the above
issues in a systematic manner within a well-motivated and complete
framework - that provided from string theory. To be clear, the
goal is \emph{not} to construct an explicit model arising from a
particular string construction. Instead, the goal is to figure out
the microscopic conditions required to address and solve the above
cosmological issues by utilizing only generic features of classes
of well-motivated low energy effective-field-theories (EFTs)
arising in various string theory compactifications. More
precisely, by ``well-motivated" it is meant that the string
compactifications stabilize the moduli as well as generate a
stable hierarchy between the electroweak and planck scales. With
developments in string compactifications in recent years, it is
now possible to realize these features in a natural manner. In
particular, in this paper we will work within the framework of low
energy supersymmetry.

The results obtained from such an analysis are quite encouraging
and can be summarized simply as follows. Within classes of string
compactifications with a mechanism of generating a stable
hierarchy and stabilizing the moduli at the same time, there is
generically a light modulus (moduli) whose mass is comparable to the
gravitino mass scale and generically much smaller than the Hubble parameter during 
inflation ($H_{inf}$). The modulus is therefore typically displaced from its minimum, 
dominates the energy density of the Universe for a long time and its decay (close to BBN)
generates a large entropy which greatly dilutes any pre-existing
baryon asymmetry. Therefore, existing mechanisms of baryogenesis
have to take this feature into account. We find, in particular,
that the mechanism of Affleck-Dine leptogenesis (via the $LH_u$
flat-direction) can still generate the observed baryon asymmetry
as well as light neutrino masses of the Universe shortly after the
end of inflation in classes of string constructions with a minimal
extension of the MSSM below the unification scale - just
consisting of right-handed neutrinos, provided certain microscopic
criteria are satisfied. These are as follows. The spectra should
consist of a gauged $U(1)_{B-L}$ symmetry with right handed
neutrinos in addition to that of the MSSM (and possible
vector-like exotics). The gauged $U(1)_{B-L}$ symmetry must be
such that the $U(1)_{B-L}$ gauge boson gets a Stuckelberg mass.
String instantons with the appropriate zero-mode structure must be
present in order to give rise to the appropriate Majorana or
Weinberg operators generating neutrino masses. Because majorana
masses are generated by instantons and are exponentially
suppressed relative to the $B-L$ breaking scale, it is natural for
the right-handed neutrinos to be hierarchical and the heaviest
right-handed neutrino to not be displaced from its minimum during
inflation, which is crucial for generating a non-zero baryon
number. Furthermore, generating the correct amount of baryon
asymmetry requires that the neutrino yukawa couplings be very
small and that the lightest left-handed neutrino is virtually
massless\footnote{It has an extremely small mass $\sim 10^{-16}
$eV.}. In an ordinary effective theory, a very small yukawa
coupling and lightest neutrino mass may not look very natural;
however since the yukawa couplings in the microscopic
constructions we are interested in are exponentially suppressed,
it is natural for them to be small. The mechanism of generating
the baryon asymmetry is similar to the ones described in
\cite{Fujii:2001dn,Fujii:2001zr} where one also requires a very
light lightest neutrino (although not as small as required here).
However, in this work we have embedded the mechanism (with subtle
differences in details) in a complete framework well-motivated
from string theory.

The above framework can also naturally solve the moduli
(gravitino) problem in both gravity mediation and gauge mediation,
although in a different manner. Within gravity mediation, there is
no moduli (gravitino) problem if the lightest modulus is of ${\cal
O}(m_{3/2}) \gtrsim 10$ TeV \footnote{It is still possible to
obtain ${\cal O}(100)$ GeV superpartner spectra within both
gravity and gauge mediation so as to be interesting at the LHC.}.
Within gauge mediation, the lightest ``modulus"(scalar field) is
typically again of ${\cal O}(m_{3/2})$\footnote{within a few
orders of magnitude.} although $m_{3/2}$ is much smaller ($\leq$
GeV) and stable. So, the main constraint in this case comes from
the decay of this modulus to the gravitino. The dark matter (DM)
candidate, be it the LSP or the gravitino, is dominantly produced
non-thermally within this framework. Therefore, the observed upper
bound on the relic density provides an important constraint on the
moduli-matter and the moduli-gravitino couplings arising within
this framework. Finally, although not a definite prediction of the
above framework, the framework allows an interesting possibility
at the LHC - that of a right-handed neutrino at the TeV scale, but
without the presence of an additional $U(1)$ gauge boson at that
scale. This could give rise to interesting predictions for the LHC
and Dark Matter (DM) and could in principle be distinguished from
scenarios with TeV scale right-handed neutrinos in which an
additional $U(1)$ gauge boson at around the same scale is also
present.

To our knowledge, this is the first time that a well defined
microscopic framework has been outlined to address all
cosmological issues starting from the end of inflation to the
beginning of BBN in a manner so as to be consistent with low
energy supersymmetry and giving rise to interesting physics at the
LHC. Each of the above microscopic conditions are naturally
compatible with each other and have been shown to be true
separately within large classes of constructions. Therefore, we
expect that there should exist a reasonably large class of
constructions within the sub-landscape of ``realistic" string
theory vacua in which all of them are satisfied
\emph{simultaneously}, leading to a successful solution of all the
above cosmological issues in a natural manner.

The outline of the paper is as follows. Section
\ref{summarystringcompact} explains the framework of string theory
compactifications which are well-motivated from a microscopic
viewpoint and also have many desirable features from the
standpoint of low-energy physics. The features of the framework
concerning the moduli and matter spectra which are most relevant
for addressing the above mentioned cosmological issues are
described in some detail. Section \ref{detailedcosmo} is a
detailed discussion of cosmology within the framework, in
particular the mechanisms which generate the baryon asymmetry in
the presence of moduli and also addresses the moduli problem and
generation of adequate amount of Dark Matter. The details of the
microscopic structure relevant for the framework, as well as the
constraints on the microscopic parameters to get the correct
baryon asymmetry, are discussed in section \ref{micro}. Section
\ref{other} briefly discusses the realization of the framework
studied in other corners of string/$M$ theory, in particular the
low energy limit of $M$ theory compactifications. In section
\ref{possibleLHCconnection}, some broad potential consequences for physical
observables are outlined. We conclude
in section \ref{conclude} and discuss future directions. In
appendix \ref{gaugemodulus}, the mass scale of the lightest
modulus is estimated within gauge mediation. In appendix
\ref{fateofN3}, it is shown that it is not possible to generate a
baryon asymmetry by affleck-dine leptogenesis if the heaviest
right-handed sneutrino is displaced from its minimum after
inflation. Appendix \ref{techdetails} deals with some technical
details concerning the computation of the lepton number shortly
after inflation, and appendix \ref{dtermmasses} estimates the
$D$-term contribution to the masses of the k\"{a}hler moduli
appearing in the various $D$-terms.

\section{Framework of ``Realistic" String
Compactifications}\label{summarystringcompact}

In this section, we give a brief and not very technical review of
relevant aspects of string compactifications giving rise to vacua
with many desirable low energy features. This will be helpful for
setting the stage in which all the above cosmological issues can
be addressed systematically. In particular, we will be interested
in the spectra of moduli in realistic string compactifications.
This will be crucial for the cosmological evolution of the
Universe after inflation. The reader primarily interested in
cosmology may skip this section, nevertheless it will be useful to
remember that in realistic string compactifications with low energy supersymmetry, 
there generically exists a light modulus with mass comparable to $m_{3/2}$, which is 
typically much smaller than the Hubble parameter during inflation ($H_{inf}$).

The most important challenges to constructing a low energy theory
arising from a string compactification are related to dynamical
issues such as moduli stabilization, supersymmetry breaking and
explaining the Hierarchy between the Electroweak and Planck
scales. Successfully addressing these opens the possibility to
construct models of particle physics beyond the Standard-Model
(SM) within the framework of string theory and study them to the
extent that testable predictions for real observables in particle
physics and cosmology can be made. However, in carrying out this
program in string theory, it has to be kept in mind that the
properties of beyond-the-Standard-Model (BSM) particle physics
models are intimately connected to the dynamical issues mentioned
above. This is because the masses and couplings of the particle
physics models depend on the properties of the vacuum (or class of
vacua) of the underlying string theory compactifications, in
particular, the values of the moduli in the vacuum. In recent
years, substantial progress has been made in the past few years
towards addressing the above dynamical issues within various
corners of the entire $M$ theory landscape, see
\cite{Dasgupta:1999ss,Balasubramanian:2005zx,Acharya:2002kv,Acharya:2006ia}.
For definiteness, we will consider Type IIB string
compactifications on Calabi-Yau orientifolds where the results are
best understood. However, the arguments given below are quite
general and only depend on certain qualitative features. Hence,
these could be generalized to many other known classes of
compactifications.

\subsection{Moduli Spectra and Low-energy Supersymmetry}\label{modulispectra}

We are interested in estimating the spectra of moduli masses in
``realistic" string compactifications where both the moduli and
the Hierarchy are stabilized at the same time. For concreteness,
we will focus on string compactifications with low energy
supersymmetry. Regarding the unification of gauge couplings within
the MSSM at $M_{GUT} \sim 10^{16}$ GeV as an important clue, in
our analysis we restrict to string compactifications with a high
compactification and string scale ($M_{GUT}\approx M_{KK}\lesssim
M_s \sim 10^{17}$ GeV)\footnote{The discrepancy of an order of
magnitude between $M_s$ and $M_{KK}$ can be naturally explained by
threshold corrections, so we will take all of them to be roughly
of the same scale.}. Another reason for such a restriction is that $M_{s} \ll M_{GUT}$, 
or equivalently, a very large compactification volume ${\cal V}$, gives rise 
to significantly more serious problems with BBN as will be seen later.

By low-energy supersymmetry it is
meant that the scale of superpartners has to be of
$\mathcal{O}$(TeV) in order to stabilize the higgs mass. String
compactification frameworks with moduli stabilization \emph{and}
supersymmetry breaking leading to low energy supersymmetry are of
the following type: supersymmetry is broken in some hidden sector
by a combination of matter and moduli fields and mediated to the
visible sector predominantly by exchange of closed strings
(gravity mediation) or open strings (gauge mediation). What
dominates depends on the separation ($d$) between the hidden
sector (which breaks supersymmetry) and the visible sector, compared
to the string length ($l_s$). In cases in which moduli
stabilization is best understood, $d \gg l_s$ which gives rise to
gravity mediation. Therefore, we will discuss gravity mediation in more detail, although
we will also comment on gauge mediation models within string theory later.

Intuitively, the above claim can be understood as follows. In
models with a vanishing (tiny) cosmological constant, the
gravitino mass in $\mathcal{N}$=1 supergravity can be written as:
\ba \label{heuristic} m_{3/2} = e^{K/2}\,\frac{W}{m_p^2} \approx
\frac{\sqrt{\sum_i F^iF_i}}{m_p} \ea where $F^i$ correspond to
fields which have non-zero $F$-term \emph{vevs}. In order to explain 
the Hierarchy, the gravitino mass $m_{3/2}$ has to vastly suppressed 
relative to $m_p$, implying that $F_i$ has to be suppressed relative to $m_p^2$. 
This implies from above that $W$ (which depends on moduli) must also be
suppressed\footnote{assuming that $e^{K/2}$ does not give a huge
suppression, which is true for compactifications with a large
string scale $M_s \gtrsim M_{GUT}$.} relative to $m_p^3$. A
natural way to obtain a small $W$ is by an exponential
suppression\footnote{there could be a small constant piece in
addition as well.}. In a generic situation, the curvature of the scalar potential 
at the minimum which determines the mass of the modulus appearing in the
exponential, is of $\mathcal{O}(\frac{W^2}{m_p^4})$, or equivalently of ${\cal
O}(m_{3/2}^2)$ from (\ref{heuristic}). The precise value, however, depends 
on details. We will carry out a detailed version of this simple argument within Type IIB
compactifications. For completeness, we will estimate the masses
of other moduli as well.

%For future purposes, it is helpful to have the following picture
%of the framework of Type IIB warped compactifications with fluxes
%and D-branes (potentially also anti D-branes\footnote{These are
%the \emph{``anti-particles"} of the D-branes.}) - The presence of
%fluxes stabilizes the dilaton, complex structure and open string
%moduli and forces the anti D-branes to settle at the bottom of the
%warped throat. Non-perturbative effects stabilize the K\"{a}hler
%moduli. Supersymmetry can be broken either by anti D-branes or by
%an appropriately realized dynamical supersymmetry breaking (DSB)
%sector at the bottom of the throat. This also provides a positive
%contribution to the scalar potential. The presence of a large
%number of fluxes can be used to tune the cosmological constant[].
%Also, supersymmetry breaking at the end of the throat is mediated
%to the visible sector residing in the bulk by gravity (moduli).
%Even though supersymmetry breaking is mediated by gravity, the
%problem of flavor changing neutral currents could be naturally and
%elegantly solved if the supersymmetry breaking sector is
%sequestered from the visible sector []. From the AdS-CFT
%correspondence, the geometrical sequestering has a nice dual in
%field theory (conformal sequestering). For further details about
%various aspects of this framework, refer to [][]. The cartoon in
%Figure \ref{cartoon} is a useful illustration. For different
%string compactifications, this picture may be different.

%\begin{figure}
%\center \epsfig{file=cartoon.eps,height=12cm, angle=0}
%\caption{cartoon} \label{cartoon}
%\end{figure}

Any string compactification preserving $\mathcal{N}$=1 SUSY in
four dimensions can be written at low energies in terms of
$\mathcal{N}$=1, $D$=4 SUGRA, which at the two-derivative level is
completely specified by a K\"{a}hler potential, superpotential and
gauge kinetic function. For IIB compactifications, these are given
by: \ba\label{KWf} K &=& -2\log(\mathcal{V}(T_i,V_i)) - \log(i\int
\Omega \wedge \Omega(U_i)) - \log(S+\bar{S}) -
\hat{K}(Y_i+\bar{Y}_i) +
\tilde{K}_{\alpha\beta}\,\bar{Q}_{\alpha}Q_{\beta}
+...\\
W &=& W_{flux}+ W_{np} + W_{matter}\nonumber\\
  &=&m_p^3\left(\frac{1}{\alpha'}
  \int G_3\wedge\Omega (S, U_i)+\sum_i\,A_i(U_j,V_k)e^{-a_i(T_i+h_i(F)\,S)}\right)+\lambda\,e^{-S(T_m)}Q_{\alpha}Q_{\beta}+
  y_{\alpha\beta\gamma}(U_j,V_k)\,
  Q_{\alpha}Q_{\beta}Q_{\gamma}+...
  \nonumber\\ f_a&=&T_a+h_a(F)\,S\nonumber\ea
The k\"{a}hler potential gets contributions from the moduli and
matter fields. The contribution of matter fields can be expressed
as an expansion around the origin as seen from
(\ref{KWf})\footnote{This is because most matter fields are
supposed to have vanishing vevs.}. $\mathcal{V}$ denotes the
volume of the internal manifold in units of the string length
($l_s$) and depends on the k\"{a}hler ($T_i$) and open string
moduli ($V_i$). $\Omega$ corresponds to the unique holomorphic
three-form of the Calabi-Yau manifold and depends on the complex
structure moduli ($U_i$) while $G_3$ corresponds to a three-form
field strength present in Type IIB string theory which depends on
the dilaton ($S$). $\tilde{K}_{\bar{\alpha}\beta}$ is the
k\"{a}hler metric of the visible matter fields $Q_{\alpha}$. The
superpotential, in addition to the classical flux contribution
which depends on $S$ and $U_i$, and the non-perturbative
contribution which depends on the k\"{a}hler moduli ($T_i$) and
the dilaton, also has a matter contribution. The renormalizable
matter superpotential contains yukawa couplings which depend on
the complex structure and open string moduli and also contains
potential mass terms which depend on the k\"{a}hler moduli, as can
be seen from (\ref{KWf}). In addition, the superpotential and the
k\"{a}hler potential could also have non-renormalizable terms;
these have been suppressed above. The potential mass terms will be
crucial for the generation of majorana neutrino masses as we will
see later. Finally, the gauge kinetic function depends primarily
on the k\"{a}hler moduli; however demanding a chiral matter sector
on the world-volume of the gauge theory implies that there is also
a dependence on the dilaton which depends on certain topological
data $h_a(F)$.

The fluxes generate contributions to the energy density of the
order of the string scale $M_s \equiv \frac{1}{\sqrt{\alpha'}}$,
which is the natural scale in the problem. Taking proper account
of the weyl rescaling to Einstein frame, the masses of the moduli
stabilized by bulk and brane worldvolume fluxes (the dilaton,
complex structure and open string moduli) can be estimated as
\cite{Choi:2004sx}: \ba m_{S,U_i,V_i} \sim \frac{\alpha'}{R^3}
\equiv \frac{M_{KK}^3}{M_{s}^2}\ea Here, $M_{KK}^{-1} \equiv R$ is
the typical size of the bulk of the Calabi-Yau. These moduli are
stabilized supersymmetrically, so they have negligible $F$-term
components. Since these moduli have masses of ${\cal
O}(10^{16}$GeV), below these energies one could integrate them out
and obtain an effective constant flux superpotential $W_0$. 

The K\"{a}hler moduli are not stabilized by fluxes. However, non-perturbative effects can in 
general give rise to a dependence on these moduli and hence help in stabilizing them. 
It has been shown that some K\"{a}hler moduli ($T_i$) can be stabilized dominantly by non-perturbative effects. However, because of the chirality of the MSSM (and possible extensions), at least the K\"{a}hler modulus which measures the volume of the cycle on which the SM gauge group is supported cannot be stabilized purely by non-perturbative effects \cite{Haack:2006cy}. This is also true for the modulus $T_m$ appearing in the superpotential in (\ref{KWf}). Therefore, some K\"{a}hler moduli ($T_{\alpha}$) have to be stabilized by a combination of other effects (arising from  $D$-terms \cite{Blumenhagen:2007sm}, K\"{a}hler corrections \cite{Cicoli:2008va}, moduli trapping \cite{Greene:2007sa}, etc.) and non-perturbative effects. As shown in \cite{Conlon:2006us}, the moduli $T_i$ are stabilized supersymmetrically at leading order while the remaining ones $T_{\alpha}$ are not. This generically leads to masses for $T_i$ which are parametrically larger than $m_{3/2}$ by a factor $\sim a_i\langle T_i\rangle$. The masses of moduli which are stabilized primarily by contributions from the K\"{a}hler potential are of the same order as $m_{3/2}$ \footnote{If ${\cal V}$ is not too large, then the mass of the overall modulus 
is also of the same order as $m_{3/2}$.}, while $D$-term contributions to masses of moduli could be much larger 
than $m_{3/2}$, as estimated in appendix \ref{dtermmasses}.

The non-perturbative effects in the superpotential ($W_{np}$)
should be of the same order as $W_{0}$ to obtain a minimum if the volume ${\cal V}$ is not too large\footnote{This is true in particular if $M_{GUT}\approx M_{KK}\lesssim M_s$ as has been assumed.}. 
Thus the flux superpotential $W_{0}$ has to be suppressed, just like $W_{np}$. This can be 
naturally provided by the discrete tuning of fluxes. In fact, it is also possible to 
choose fluxes such that $W_0$ vanishes \cite{Blumenhagen:2007sm}. In the presence of warping, there are also throat moduli ($Y_i$) which are
stabilized by the fluxes, their vacuum
values related to the warp factor ($e^{A_{min}}$) at the tip of the throat. In
particular \cite{Choi:2004sx}: \ba m_{Y_i} \sim
e^{A_{min}}\,\frac{\alpha'}{R^3} =
e^{A_{min}}\,\frac{M_{KK}^3}{M_s^2}\ea 

Without additional effects, the vacuum obtained after stabilizing the moduli generically has negative vacuum energy, i.e. it is an anti de-Sitter (AdS) vacuum. Therefore, a positive contribution to the vacuum energy is needed to obtain a dS vacuum with a positive cosmological constant. Moreover, the positive contribution has to be (finely) tuned so that the cosmological constant has the observed value, for which no satisfactory dynamical solution exists at present. Various mechanisms giving rise to a positive contribution to the vacuum energy exist, such as explicit supersymmetry breaking contributions from anti D-branes, or from 
$F$-term and $D$-term uplifting by matter fields. Since our primary interest is the spectra of moduli (scalar fields), the masses of new scalar degrees of freedom which appear in the mechanisms above must also be taken into account after fine-tuning the cosmological constant. $F$-term uplifting, for example, generically gives rise to masses for these matter fields of ${\cal O}(m_{3/2})$\cite{Acharya:2006ia,Lebedev:2006qq}, while $D$-terms could give rise to large masses for the moduli, as estimated in appendix \ref{dtermmasses}. 

We would also like to comment on gauge mediation models within string theory. One could try to imagine a situation
in which \emph{all} moduli are stabilized at a high scale in an almost supersymmetric and minkowski vacuum. 
Although there do not currently exist explicit compactifications which realize this situation (however, see \cite{Diaconescu:2005pc} for some work in this direction), but one could hope that 
future developments could accomplish this. Supersymmetry could then be broken by a dynamical mechanism in a hidden matter sector and be mediated to the visible sector by gauge interactions if $d \leq l_s$. This is the philosophy of many \emph{local} models in string theory \cite{Kawano:2007ru}. However, it is important to note
that even within such gauge mediation models, there exists a modulus-like
scalar field (the scalar partner of the goldstino, or the $D$-flat direction comprising the vector-like messengers) which
generically gets a mass comparable to $m_{3/2}$ (within a
few orders of magnitude). This is argued in appendix
\ref{gaugemodulus} for generic models of gauge mediation.

It is important to note that in compactifications in which the volume ${\cal V}$ is very large (or equivalently, $M_s$ 
is much smaller than the traditional $M_{GUT}$), such as which could arise in LARGE volume compactifications \cite{Conlon:2007gk} or in the local models above if they are sufficiently decoupled from gravity, the overall volume modulus could be much lighter 
than $m_{3/2}$. Thus, these models would cause very serious problems for BBN unless mechanisms exist which could sufficiently dilute the entropy produced at late times by the decay of these extremely light moduli. Although this is possible in principle, there do not exist concrete mechanisms within string theory at present which realize it. Therefore, as mentioned earlier, we do not consider this situation. 

To summarize, realistic string compactifications with a mechanism of generating and stabilizing
the Hierarchy between the electroweak and Planck scales while
stabilizing the moduli, give rise to two sets of moduli - one very
heavy and one light. The light moduli are typically comparable to $m_{3/2}$. 
With low energy supersymmetry, the hubble parameter during inflation $H_{inf}$ is typically 
much larger than the moduli or gravitino mass, which generically destabilizes the light moduli. 
This feature is expected to be true for other classes of
string compactifications as well. For example, the above feature
is satisfied for moduli spectra in realistic $M$ theory
compactifications studied in
\cite{Acharya:2008bk}. Hence, we will assume the above spectra of moduli henceforth.

\subsection{Visible Sector Model Building}\label{modelbuilding}

Now that we understand some of the important features of the class
of vacua obtained in the above framework, the next step is to look
at fluctuations around these vacua, those pertaining to matter and
gauge degrees of freedom. This corresponds to constructing the
matter and gauge spectrum comprising a beyond-the-SM particle
physics model. As explained earlier, this is not the subject of
this paper. Nevertheless, it is worthwhile to elaborate a little
on visible sector model-building.

Most work on explicit string model-building is focussed on
computing spectra on compactifications on toroidal
orbifolds/orientifolds and Gepner models, where CFT techniques are
available. Computing spectra on a general compact Calabi-Yau is
extremely challenging. However, considerable progress has been
made in computing spectra in non-compact (local) Type II
constructions with D-branes at singularities where gravity can be
decoupled at leading order. In both cases, semi-realistic spectra
for beyond-the-SM physics have been constructed. We will not
concern ourselves with constructing specific singularities or
toroidal constructions realizing the MSSM or its extension
thereof; rather we are interested in studying generic features
which are crucial in solving the cosmological problems outlined in
the introduction and would be relevant even if a particular
explicit construction is not.

Leptogenesis provides us with an elegant mechanism of explaining
the origin of neutrino masses as well as the baryon asymmetry of
the Universe in one theoretical framework. Therefore, with the
principle of \emph{Occam's Razor} in mind, in this work we will
focus on leptogenesis as providing the mechanism for baryon
asymmetry of the Universe. The requirement of small neutrino
masses requires the presence of the lepton number violating
operator - $\frac{\kappa}{M}LH_uLH_u$ in the superpotential, with
$\kappa$ a dimensionless coupling and $M$ a large mass-scale.
There are two ways of generating this operator - a) It may be
present in the microscopic construction itself (``Weinberg" case)
and/or b) It may be generated at lower energies by integrating out
right-handed neutrinos ($N_R$) which appear in many semi-realistic
spectra arising from string constructions (`` see-saw case").
These constructions typically have an additional gauged $U(1)$ --
$U(1)_{B-L}$ to be precise. The existence of a gauged $U(1)_{B-L}$
is quite natural as it is the unique flavor independent
anomaly-free $U(1)$ asymmetry, and string theory only allows gauge
symmetries. Anomaly cancellation (including the gravitational
anomaly) in fact requires the existence of three families of
right-handed neutrinos. However, for this mechanism to work, large
majorana masses for right handed neutrinos have to be generated.
Both of the above mechanisms will be discussed within the context
of string constructions in section \ref{applyneutrinos}.

For concreteness, in the following, we will assume that the
visible sector consists of a supersymmetric standard model with a
 $U(1)_{B-L}$ gauge group in addition to that of the SM and a matter
 spectrum consisting of the MSSM, possibly vector-like exotics and
 three right-handed neutrinos.

\section{Cosmology within the Framework}\label{detailedcosmo}

In this section, we will discuss the cosmological implications of
the framework described above during and after inflation until the
beginning of BBN. As stated earlier, we will assume the existence
of an inflationary phase in the very early Universe which solves
the flatness and horizon problems and also gives rise to almost
scale invariant density perturbations as observed.

Many supersymmetric models such as the MSSM commonly have a
degenerate set of vacua at the level of renormalizable terms,
meaning that there are many directions in the space of scalar
fields where the potential vanishes classically in the
supersymmetric limit with $m_p \rightarrow \infty$. These
directions are therefore known as ``flat-directions". These
flat-directions are, however, lifted by supersymmetry breaking and
non-renormalizable terms in the superpotential \cite{Dine:1995kz}.
We will specifically be interested in the evolution of moduli,
sneutrinos and ``flat-directions" during and after inflation as it
will be crucial for the estimation of the baryon asymmetry.

\subsection{Evolution during the Inflationary Phase}\label{duringinf}

As explained in \cite{Dine:1995kz}, the finite energy density of
the Universe during inflation breaks supersymmetry. The finite
energy supersymmetry breaking is transmitted to the moduli,
sneutrinos and flat-directions by the cross-coupling of the
inflaton (which dominates the energy density during inflation by
definition) and the above fields. Ordinary hidden sector
supersymmetry breaking corrections to the scalar potential are
much smaller during inflation since $H_I \gg m_{3/2}$ in vacua
with low energy supersymmetry. Parameterizing all these couplings,
one finds the following contribution to the scalar potential for
the moduli, sneutrinos (if present in the spectrum) and flat
directions ($\phi$) during inflation \cite{Dine:1995kz}: \ba
\label{pot}V_{inf} &=&
\sum_{i=1}^N\,c_1^iH_I^2|X_i|^2+M_{X_i}^2|X_i|^2+(b_1^{i}H_IM_{X_i}X_iX_i+c.c)+....\nonumber\\
&&
+\,c_2^kH_I^2|\tilde{N}_k|^2+M_{N}^{k}|\tilde{N}_k|^2+(b_2H_IM_N^k\tilde{N}_k\tilde{N}_k+c.c)+
..\nonumber\\ && +\, c_3H_I^2|\phi|^2+(\frac{a\lambda
H_I\phi^n}{nM^{n-3}}+c.c.)+|\lambda|^2\frac{|\phi|^{2n-2}}{M^{2n-6}}+
(m_0^2|\phi|^2+\frac{Am_{3/2}\lambda\phi^n}{nM^{n-3}}+c.c)+...\ea
Here $\{c_{1}^i,c_2^k,c_3\}$, $\{b_1^{i},b_2\}$ and
$\{a,A,\lambda\}$ are model-dependent coefficients typically of
$O(1)$ but can be of either sign. From (\ref{pot}), we see that
the moduli $X_i$ and sneutrinos $\tilde{N}$ have a mass term in
the potential because the moduli are stabilized and the sneutrinos
have a supersymmetric majorana mass term. The flat directions
$\phi$ on the other hand are massless at leading order by
definition. The scalar potential depends on $\phi$ only through
supersymmetry breaking (both hubble induced and hidden sector) and
by non-renormalizable terms in the superpotential. These terms are
also present for the moduli and sneutrinos. However, the hidden
sector supersymmetry breaking and non-renormalizable terms are not
written for simplicity as they are much smaller.

We will start by studying the evolution of moduli during
inflation. In most natural models of inflation, the hubble
parameter during inflation ($H_I$) is ${\cal O}(10^{13}-10^{14})$
GeV. For our purposes, we will assume $H_I \sim 10^{12}-10^{13}$
GeV for our solutions to be consistent conservatively, as will be
seen later. For larger values of $H_I$, it is still possible for
our solutions to be consistent but less parameter space is
available. As stated earlier, the coefficients $c_1^i$ depend on
the concrete model of inflation and can be of either sign. If they
turn out to be positive and O(1), then from (\ref{pot}) it is
clear that the effective mass-squared parameter for all the moduli
$(M^{eff}_{X_i})^2 = M_{{X_i}}^2+c_{1}^iH_I^2$ is positive and
$\gtrsim {\cal O}(H_I^2)$. However, a negative sign for $c_1^i$ is
possible for non-minimal couplings between the inflaton and the
relevant field, which is in fact a generic possibility within
string theory. In this case, the effective mass-squared parameter
for the moduli are given by: \ba (M^{eff}_{X_i})^2 =
M_{X_i}^2-|c_1^i|H_I^2\ea As explained in section
\ref{modulispectra}, the complex structure ($U_i$), dilaton ($S$)
and open string moduli ($V_i$) naturally obtain masses of ${\cal
O}(10^{16})$ GeV. Therefore, the effective mass-squared parameter
for these moduli $(M^{eff}_{X_{heavy}})^2$ is still positive and
they settle down to the true (late time) minima in about a hubble
time. They are thus not displaced from their true minima during
inflation. The masses of the throat moduli ($Y_i$) are ${\cal
O}(10^{11}-10^{12})$ GeV. So their fate depends more on the
concrete model of inflation and the magnitude of the $c_1^{Y_i}$
in particular. For negative $c_1^i$ slightly small in magnitude
for some reason ($\lesssim 0.1$), the moduli $Y_i$ will not be
displaced, otherwise they will. Finally moving on to the light
moduli, which we denote by $X_{light}$ in general, we see from
section \ref{modulispectra} that they have masses of ${\cal
O}(m_{3/2})$, i.e. of $\gtrsim 10$ TeV \footnote{in order to solve
the moduli problem, see section \ref{moduliDMproblems}.}.
Therefore, these moduli will generically be displaced from their
true minima for a wide range of $c_1^i$. As will be shown later,
our final conclusions regarding the baryon asymmetry and dark
matter will not change whether or not the moduli $Y_i$ are
displaced, as long as the light moduli $X_{light}$ \emph{are}
displaced during inflation. Therefore, for simplicity and
concreteness, it will be assumed that all the heavy moduli
($m_{X_{heavy}} \gg m_{3/2}$) settle in their true minima and the
light moduli $X_{light}$ are displaced from their true minima by a
large amount.

The situation with sneutrinos is quite interesting. As explained
in section \ref{modelbuilding}, many BSM models for new physics
arising from string constructions have a $U(1)_{B-L}$ symmetry
with three families of right-handed neutrinos (and sneutrinos in a
supersymmetric model). In such constructions, for the right handed
neutrinos and sneutrinos to receive a majorana mass term, the
$U(1)_{B-L}$ symmetry needs to be broken. However, until recently
it has proven difficult to break the above symmetry in the desired
manner without not producing other undesirable effects such as
dangerous $B$ and $L$ violating operators at the same time. It was
recently shown in \cite{Blumenhagen:2006xt} that under certain
conditions, the perturbatively forbidden majorana mass term could
be generated in a natural way by stringy non-perturbative effects
(see section \ref{micro} for details). These non-perturbative
effects arise from euclidean brane instantons; hence the majorana
masses generated from these effects are exponentially suppressed.
The scale at which the gauged $U(1)_{B-L}$ is broken is roughly
the string scale, giving rise to a mass for the $B-L$ gauge boson
of ${\cal O}(M_s)$. This gives rise to a mass term in the
superpotential for the right-handed neutrinos (the third term in
the expression for the superpotential in (\ref{KWf})). The
majorana masses are given by: \ba\label{massrh} M_N^{ab} \equiv
\lambda\,v_{B-L} =
e^{K/2}\,\sum_r\,d^r_ad^r_b\,e^{-S_{E3}^{(M),r}(T_m)}\,m_p\ea
where $\lambda$ is a dimensionless coefficient, $d^r_a,d^r_b$ are
constants generically of ${\cal O}(1)$, and $S_{E3}^{(M),r}$, the
action of the $r^{th}$ instanton, is (classically) equal to the
world-volume of the instanton. For details, refer to section
\ref{applyneutrinos}. To generate majorana masses for three
right-handed neutrinos, three or more contributing instantons are
required. Since the world-volume of each instanton will in general
differ by ${\cal O}(1)$, the flavor structure above implies that
there will generically be a hierarchy among the right handed
majorana mass eigenvalues. This implies that it is very natural
for the \emph{heaviest} right-handed sneutrino $\tilde{N}_3$ to
have a mass greater than $H_I\,(\sim 10^{12-13}$ GeV), implying
(from arguments in the previous paragraph) that it is not
displaced from its true (late-time) minimum during inflation. The
other sneutrinos may or may not be displaced depending on their
masses and the magnitude and sign of the coefficient $c_2^k$.
However, the desired baryon asymmetry can only be generated if the
heaviest right-handed sneutrino does not get displaced during
inflation. Therefore, we will only focus on the heaviest right
handed sneutrino and assume that it is not displaced from its true
minimum.

Since the heaviest right handed sneutrino ($\tilde{N}_3$) is
naturally expected to settle to its minimum during inflation as
argued above, one can integrate out $\tilde{N}_3$ and consider the
corresponding $LH_u$ direction (giving the \emph{lightest}
left-handed neutrino mass), which becomes a flat-direction as it
only gets contributions from supersymmetry breaking and
non-renormalizable operators. The lowest dimension operator
involving the $LH_u$ flat direction is $\frac{\kappa\,
LH_uLH_u}{M}$ as mentioned in section
\ref{modelbuilding}\footnote{This could arise by itself
(``weinberg" case) and/or by integrating out right-handed
neutrinos (``see saw" case).}. It is worth noting, that even
though this operator violates $B-L$, it does preserve $R$-parity.
We now turn to the evolution of flat-directions during inflation,
focussing on the $LH_u$ direction, also denoted by $\phi$, in
particular.

The fate of flat directions during inflation also depends on the
coefficient of the Hubble induced mass term ($c_3$). However,
since the flat-directions have gauge and yukawa couplings to other
matter fields, one also needs to take into account the effect of
renormalization (RGE) from the scale at which they are introduced
(typically the compactification scale in string compactifications
$\sim M_{GUT},\, M_s$) to lower scales ($\sim H_I$). So, it could
happen that RG effects could make the effective mass-squared
parameter $(m^{eff}_{\phi})^2$ negative at some scale $Q_c$ even
if $c_3$ is positive ${\cal O}(1)$ \cite{Enqvist:2003gh}. As
argued in \cite{Enqvist:2003gh}, the $LH_u$ direction is most
likely to develop a negative mass-squared parameter for a wide
range of $c_3$ (both positive and negative) due to RG evolution
because of the large top yukawa coupling. The $LH_u$ direction has
other advantages as well in addition to providing a mechanism for
generating the $B-L$ asymmetry and giving rise to viable neutrino
masses. Once the $LH_u$ direction is displaced, most
flat-directions (in the MSSM) are lifted at the renormalizable
level so that only the $LH_u$ direction needs to be considered for
further analysis. Finally, the $LH_u$ direction is free from the
$Q$-ball problem, to be discussed in section \ref{allconstraints}.
Therefore, the $LH_u$ flat-direction is the most natural and
robust candidate for generation of the baryon asymmetry.

To summarize, the heavy moduli, the heaviest right handed
sneutrino and most flat-directions settle down in their true
minima very quickly while the light moduli and the $LH_u$ flat
direction (corresponding to the lightest left-handed neutrino) are
displaced from their minima by a large amount of ${\cal O}(H_I)$.
For the light moduli, naively one might think that their
post-inflationary evolution is fraught with the overshoot problem
\cite{Brustein:1992nk}. However, as was argued in
\cite{Acharya:2008bk,Kaloper:1991mq,Conlon:2008cj}, in the
presence of matter or radiation (which is true in the present
case) the fields can be easily guided towards the global minimum
without overshooting. Thus, there is no overshoot problem in this
framework. We now study the post-inflationary evolution of the
$LH_u$ flat-direction in detail.

\subsection{Post-Inflationary Evolution}\label{postinf}

Let us study the potential (\ref{pot}) in more detail, with
particular attention to the $LH_u$ direction and $\tilde{N}_3$.
The analysis of this section is similar to the one in
\cite{Fujii:2001dn,Fujii:2001zr} which was the first to look at
the evolution of the $LH_u$ flat direction with a gauged
$U(1)_{B-L}$ in a systematic manner. However, there are some
important differences. The origin of the $U(1)_{B-L}$ breaking and
that of the mass of the $U(1)_{B-L}$ gauge boson is different
leading to some subtle differences in the analysis; the generation
of left-handed and right-handed neutrinos uses a different
mechanism leading to different mass-scales; and finally, the
embedding of the baryon asymmetry generation mechanism in a
complete string framework gives rise to different regions of
allowed parameter space relevant for neutrino masses. In the
following, we will only outline the important steps here and leave
a detailed analysis to appendix \ref{techdetails}.

The scalar potential for these fields can be schematically written
as: \ba \label{schematic}V &=&
V_{susy}+V_{D}+V_{hubble}+V_{soft}\ea The (normalized)
superpotential for the canonically normalized $\hat{L}, \hat{H}_u$
and $\hat{N}_3$ is given by:\ba \label{W1}\hat{W} =
\hat{h}\hat{N}_3\hat{L}\hat{H}_u+\hat{M}_N\,\hat{N}_3\hat{N}_3+W'(\psi_i,\hat{L},\hat{H}_u,\hat{N}_3)\ea
$\psi_i$ stand for fields charged under $U(1)_{B-L}$ (in addition
to the MSSM and right handed neutrino fields) with charge $q_i$.
$W'$ contains terms depending on $\psi_i$ alone as well as terms
containing cross-couplings between $\psi_i$ and
$\{\hat{L},\hat{H}_u,\hat{N}_3\}$. The mass parameter for
$\hat{N}_3$ arises when the $B-L$ gauge symmetry is broken. Since
$B-L$ is broken at $v_{B-L} \sim M_s$, one can write $\hat{M}_N$
as $\hat{M}_{N} = \lambda\,v_{B-L} \sim \lambda\,M_s$ with
$\lambda$ a dimensionless coefficient. In the presence of a
$U(1)_{B-L}$ gauge symmetry, a non-trivial contribution from the
$U(1)_{B-L}$ $D$-term arises, and is given by: \ba
\label{Dterm}V_D =
\frac{g_{B-L}^2}{2}\,(|\hat{\tilde{N}}|^2-|\hat{\tilde{L}}|^2+\sum_i\,q_i\psi_i\partial_i
\,K-\langle \frac{1}{4\pi^2}\partial_{T_m}K \rangle)^2\ea where
$K$ is the K\"{a}hler potential. The last term in (\ref{Dterm})
depends on $T_m$ and behaves as an effective Fayet-Iliopoulos (FI)
parameter once $T_m$ gets a {\it vev} (its origin will be
explained in section \ref{micro}). $T_m$ is precisely the modulus
which appears in the mass parameter for $\hat{\tilde{N}}_3$ (see
(\ref{massrh})). It is assumed that $T_m$ is stabilized by a
combination of effects such as higher order corrections to the
k\"{a}hler potential and/or moduli trapping. As explained in the
previous section, the mass of the heaviest right handed sneutrino
($\hat{M}_N$) can naturally be greater than the hubble parameter
during inflation ($H_I$). Thus, $\hat{\tilde{N}}_3$ quickly
settles down to its minimum during inflation. The minimum is
approximately given by: \ba \langle \hat{\tilde{N_3}} \rangle
\approx -{\cal
O}(1)\,\frac{\hat{h}\,\hat{\tilde{L}}\hat{{H_u}}}{\hat{M}_{N}}\ea
The above expression for $\hat{\tilde{N}}_3$ can be substituted in
(\ref{schematic}) to generate a potential for the $LH_u$ flat
direction, obtained by substituting the neutral components of the
$\tilde{L}$ and $H_u$ fields with $\phi$. One might worry that the
large $D$-term contribution in (\ref{Dterm}) will destroy the
flatness of $\phi$. However, the fields $\psi_i$ in (\ref{Dterm})
will shift in general to make the $D$-term vanish and minimize the
potential. This is justified since the curvature around the
$D$-term potential is of ${\cal O}(M_s^2)$ which is much larger
than that of the $F$-term potential. This has been shown
explicitly in appendix \ref{techdetails}. Thus, $\phi$ can remain
approximately flat and obtain a large expectation value during
inflation. After integrating out $\hat{\tilde{N}}_3$ and other
fields like $\psi_i,\bar{\psi}_i$, one gets the following
expression for the potential for $\phi$: \ba \label{potphi}V(\phi)
&=&
V_0-c_{\phi}^2\,H^2|\phi|^2+k_{\phi}^2\,H^2\frac{|\phi|^4}{M_s^2}+{\cal
O}(1)\, \frac{H^2|\hat{h}|^2|\phi|^4}{|\lambda|^2\,M_s^2}- {\cal
O}(1)(a\,H\,\frac{\hat{h}^2{\phi}^4}{\lambda\,M_s}+h.c)+{\cal
O}(1)\, \frac{|\hat{h}|^4|\phi|^6}{|\lambda|^2\,M_s^2}...\nonumber\\
& &
+(m_{\phi}^2|\phi|^2+\frac{\hat{h}^{2(n-3)}\,A\,m_{3/2}\gamma\phi^n}{n\,\lambda^{n-3}\,M_s^{n-3}}+c.c)+...
\ea  where $V_0$ is the $\phi$-independent contribution,
$\{c_{\phi}^2,k_{\phi}^2,a,A\}$ are dimensionless coefficients
typically of ${\cal O}(1)$. ``..." stands for higher order terms
proportional to powers of $(\frac{\hat{h}}{\lambda})$ and
$(\frac{\hat{h}^2}{\lambda})$, among others. We consider the case
in which $(\frac{\hat{h}}{\lambda})$ and
$(\frac{\hat{h}^2}{\lambda})$ are suppressed so that it is
possible to neglect those terms. In section \ref{applyneutrinos},
it will be argued that these can be naturally achieved. We will
also see later that in order to produce the required baryon
asymmetry, one needs a very small value of
$(\frac{\hat{h}^2}{\lambda})$. Thus, terms proportional to
$(\frac{\hat{h}^2}{\lambda})$ are required to be suppressed for
consistency.

It is worthwhile to understand the origin of the various terms in
(\ref{potphi}). The potential for $\phi$ arises from supersymmetry
breaking (both hubble induced and hidden sector) and
non-renormalizable terms (after integrating out
$\hat{\tilde{N}}_3$ and $\psi_i,\bar{\psi}_i$). The terms in the
second line in (\ref{potphi}) are contributions from hidden sector
(soft) supersymmetry breaking which are much smaller than the
hubble parameter during and just after inflation\footnote{As
mentioned earlier, we are considering ``natural" high scale
inflation with low scale supersymmetry.}. Under the above
conditions, the minimization of the potential (\ref{potphi}) with
respect to $|\phi|$ can be simplified:
\begin{eqnarray} \frac{\partial\,V}{\partial\,|\phi|} &\approx&
-2c_{\phi}^2H^2|\phi|+4k^2_{\phi}\,H^2\frac{|\phi|^3}{M_s^2} = 0
\nonumber\\
\implies {|\phi|}^2 &\approx&
\frac{c_{\phi}^2}{2k^2_{\phi}}\,M_s^2\end{eqnarray} Since
$c_{\phi},k_{\phi}$ are of ${\cal O}(1)$ (see appendix
\ref{techdetails}), one has\footnote{One needs $|\phi| \lesssim
M_s$ for consistency of the solution, see appendix
\ref{techdetails}.}: \ba \label{vev1}|\phi| \approx {\cal
O}(1)\,M_s\ea Even though terms proportional to
$(\frac{\hat{h}^2}{\lambda})$ in (\ref{potphi}) are suppressed, as
the hubble parameter decreases after inflation, eventually the
fourth and fifth terms (proportional to
$(\frac{\hat{h}^2}{\lambda})$) in (\ref{potphi}) will become
comparable to the second and third terms. This will happen when :
\ba H \approx {\cal O}(1)\,\frac{\hat{h}^2}{\lambda}\,M_s \equiv
H_0\ea Remember that $M_s$ is greater than $H_I$. Thus, for
solution (\ref{vev1}) to be consistent, one must have $M_s > H_{I}
> H
> H_0$. This is possible since we have $\frac{h^2}{\lambda} \ll 1$. It will be
argued in section that values of $\frac{h^2}{\lambda} \ll 1$ are
quite natural in string constructions where the majorana mass of
neutrinos arises from string instantons. In the regime $M_s \sim
v_{B-L} > H_{I} > H > H_0$, the dominant contribution to the
potential for the phase of $\phi$ ($\phi = |\phi|\,e^{i\theta}$)
in (\ref{potphi}) is given by:\ba V(\hat{\theta})
\approx m_{\hat{\theta}}^2\,|\hat{\theta}|^2;\;\;\hat{\theta}\equiv |\phi|\theta\nonumber\\
{\rm where}\;\;\; m_{\hat{\theta}}^2 \approx \frac{H_I}{M}|\phi|^2
\approx {\cal O}(1)\frac{\hat{h}^2}{\lambda}\,M_s\,H < H^2 \ea
Thus, $\hat{\theta}$ is not settled at the minimum of its
potential during inflation and just after inflation. When the
hubble parameter drops to $\sim H_0$ such that $m_{\hat{\theta}}^2
\sim H_0^2$, the hubble $A$-term in the potential can kick $\phi$
in the phase direction providing a torque. This is crucial for
generating the baryon asymmetry, as we will now argue.

\subsubsection{Generation of Lepton Number}\label{lepnum}

After the end of inflation, the Universe is dominated by the
coherent oscillations of the inflaton making it matter dominated.
This means that the scale factor $R$ goes like $H^{-2/3}$ and $H$
keeps decreasing after inflation. We will assume that the
production of lepton number takes place in this epoch, justifying
it in the next subsection \ref{allconstraints}.

After inflation, in the regime $v_{B-L} > H_{I} > H > H_0$,
$|\phi| \approx v_{B-L} \sim M_s$  as shown above. The phase of
$\phi$ is displaced from its minimum during this regime. When $H$
drops to values such that $H \sim H_0$, the curvature of the
potential along the phase direction becomes comparable to $H$, and
thus the hubble $A$-term provides a torque to the phase of $\phi$.
After this time,
$(\frac{\hat{h}^2}{\lambda})^2\,\frac{|\phi|^6}{4M_s^2}
>
\{k_{\phi}^2\,H^2\frac{|\phi|^4}{M_s^2},|a|H\frac{\hat{h}^2}{\lambda}\frac{|\phi|^4}{M_s}\,\sin{(arg(a)+4arg(\phi))}\}$
This implies that $|\phi| \approx \sqrt{MH}$ for $H_0>H>H_{osc}$
where $M \equiv \frac{\lambda\,M_s}{\hat{h}^2}$. Here, $H_{osc}$
corresponds to the time when $|\phi|$ starts oscillating about its
true minimum. The value of $H_{osc}$ is determined by a
combination of susy breaking effects and thermal effects (more on
this in the next subsection \ref{allconstraints}). As shown in
\cite{Fujii:2001dn}, $|\phi|$ drops as $H^{\alpha}$ ($\alpha
\gtrsim 1$) for $H \lesssim H_{osc}$.

The lepton number density ($n_L$) related to $\phi$ is given by:
\ba n_L = \frac{1}{2}i(\dot{{\phi}^*}{\phi}-{\phi}^*\dot{\phi})\ea
 As already explained above, for
$v_{B-L} > H_{I} > H > H_0$, the phase of $\phi$ is displaced from
its minimum and no lepton number is generated. For $H_0>H$, the
time evolution of $n_L$ in the expanding universe for potential
(\ref{potphi}) is given by: \ba\label{nL1} \dot{n_L}+3Hn_L \approx
\frac{H}{M}\mathrm{Im}(a\phi^4)+\frac{m_{\phi}}{4M}{\rm
Im}(A\lambda\phi^4) +...\ea The total lepton number $N_L\equiv
R^3\,n_L$ is therefore given by: \ba\label{nL2} N_L(t) \approx
\int^t
dt\,R^3\left[\frac{2H}{M}\,|a||\phi|^4\,\sin{(arg(a)+4arg(\phi))}+\frac{m_{\phi}}{2M}|A||\lambda||\phi^4|
\sin{(arg(A)+arg(\lambda)+4arg(\phi))}\right]\ea For $H_0 > H
> H_{osc}$, one has: \ba \frac{R^3H|a||\phi|^4}{M} \sim |a|\,MH \sim
\frac{|a|M}{t}\nonumber\\ \frac{R^3m_{\phi}|A\lambda||\phi|^4}{M}
\sim |A\lambda|\, m_{\phi}M\ea Also, the argument of the sine
function of both terms in (\ref{nL2}) oscillates with a frequency
$ f \sim H$, since $m_{\hat{\theta}} \sim H$ in this regime. Thus,
the lepton number at $H \sim H_{osc}$ is given by: \ba
\label{totalnL}N_L(t\sim H_{osc}^{-1}) \sim {\cal
O}(1)\left(M\log{(\frac{H_0}{H_{osc}})}+
(m_{\phi}\,M)(H_{osc}^{-1}-H_0^{-1})\right)\ea For later times
($H_{osc}>H$), assuming the most conservative case that $|\phi|$
damps as $H^{\alpha}$ with $\alpha\approx1$, we have: \ba
\frac{R^3H|a||\phi|^4}{M} \sim
\frac{H^3}{M} \sim \frac{1}{M\,t^3}\nonumber\\
\frac{R^3m_{\phi}|A\lambda||\phi|^4}{M} \sim \frac{H^2m_{\phi}}{M}
\sim \frac{m_{\phi}}{M\,t^2}, \ea implying that lepton number
production is strongly suppressed for $H<H_{osc}$. This means that
the total lepton number is fixed when $H \sim H_{osc}$, giving
rise to (\ref{totalnL}) for the total lepton number.

\subsubsection{Satisfying All Constraints}\label{allconstraints}

The above result is quite interesting; however  some implicit
conditions need to be satisfied for it to hold. First, early
oscillations of $\phi$ have to be avoided so that the lepton
asymmetry is not suppressed (this has been implicitly assumed in
the previous subsection. Also, as assumed one has to show that the
production of lepton asymmetry takes place in the inflaton
matter-dominated era, i.e. before the reheating process is
completed.

The above constraints can be encapsulated in the following set of
conditions: \ba H_0 > H_{osc} \gtrsim \Gamma_{inf} \ea $H_{osc}$
is determined by a combination of thermal effects and
supersymmetry breaking effects, as explained in
\cite{Fujii:2001zr}. The dominant supersymmetry breaking effect is
just given by the soft mass term for $\phi$ -- $m_{\phi} \sim
m_{3/2}$. Also, although we have assumed the energy density is
dominated by the inflaton during the inflaton oscillation era,
there is still a dilute plasma in this regime with a temperature
given by: \ba\label{plasmaT} T = [(T^{inf}_R)^2m_pH]^{1/4}\ea
where $T^{inf}_R$ is the reheat temperature after inflation. This
dilute plasma gives rise to two classes of thermal effects -- a) A
thermal mass term for $\phi$ ($(m^{th}_{\phi})^2\sim c_k f_k^2
T^2$) is induced when the fields which couple to $\phi$ have an
effective mass $f_k|\phi| < T$, giving rise to a potential
contribution $V_1 \sim (m^{th}_{\phi})^2|\phi|^2$. b) Another
thermal contribution to the potential arises because the $SU(3)$
gauge symmetry remains unbroken along the $LH_u$ direction and
down type (s)quarks are also massless along this direction. This
gives rise to an effective potential $V \propto g_s^2T^4$.
However, because the RG evolution of $g_s$ depends on the
effective masses ($\sim f_k|\phi|$) of fields $\psi_k$ which are
coupled to $\phi$, this implies a potential contribution $V_2 \sim
\alpha_s\,T^4\left[\sum_{y_u|\phi|>T}\frac{2}{3}T(R_u)\right]\log{\left(\frac{|\phi|^2}{T^2}\right)}$.

After inflation the hubble parameter decreases and eventually the
negative hubble-induced mass term in the potential for $\phi$ is
surpassed by the above contributions, i.e. \ba \label{osc}H^2
\lesssim m_{\phi}^2+\sum_{f_k|\phi|<T}\,c_k f_k^2
T^2+\alpha_s^2(T)\frac{T^4}{|\phi|^2}\ea which sets the value of
$H_{osc}$. The second term in (\ref{osc}) gives rise to two
sub-cases depending on whether $f_i$ is small or large
\cite{Asaka:2000nb}. More precisely, one finds \ba\label{osc}
H_{osc} &=& max\left[m_{\phi},H_i,\alpha_s T_R^{inf}
\left(\frac{m_p}{M}\right)^{1/2}\right]\nonumber\\
{\rm where}\;\; H_i &\equiv&
min\left[\frac{m_p(T_R^{inf})^2}{f_i^4\,M^2},(c_i^2f_i^4 m_p
(T_R^{inf})^2)^{1/3}\right] \ea where
$M\equiv\frac{\lambda}{\hat{h}^2}M_s$ as before. In order to avoid
early oscillations, $H_0
> H_{osc}$ is required. From (\ref{plasmaT}) and (\ref{osc}), this
implies : \ba \label{TRinf}T^{inf}_R <
min\left[min_k\{max(\frac{f_k
M_s^{3/2}}{c_k^{1/4}m_p^{1/2}},\frac{M_s^3}{c_k f_k^2(m_p
M^3)^{1/2}})\},\frac{M_s^2}{\alpha_s(m_p M)^{1/2}}\right]\ea If
the mass parameter $M\equiv\frac{\lambda}{\hat{h}^2}M_s$ in
(\ref{TRinf}) is much larger than $M_s$, then it is clear that
$H_{osc}$ is determined by the third term in (\ref{TRinf}). For
this, $\frac{\hat{h}^2}{\lambda} \ll 1$ is required. In fact, it
will be shown in the coming sections that
$\frac{\hat{h}^2}{\lambda} \sim 10^{-12} \ll 1$ is required for
producing the correct baryon asymmetry, which can be obtained
naturally in certain classes of string constructions. Then, for
$M_s \approx 10^{17}$ GeV and $\frac{\hat{h}^2}{\lambda} \sim
10^{-12}$, the constraint (\ref{TRinf}) gives rise to $T_R^{inf}
\lesssim 10^{11-12}$ GeV. This seems to be quite natural. One also
has to check the assumption that the lepton asymmetry is produced
during the inflaton oscillation dominated era, implying the
condition $H_{osc} > \Gamma_{inf}$. Thus, $H_{osc} > \Gamma_{inf}$
requires:\ba T_R^{inf} \lesssim
\alpha_s\,\frac{m_p^{3/2}}{M^{1/2}}(\frac{90}{\pi^2 g_*})^{1/2}\ea
which is again satisfied for $T_R^{inf} \lesssim 10^{11-12}$ GeV
and $\frac{\hat{h}^2}{\lambda} \sim 10^{-12}$. Thus, we have shown
that the above constraints can be satisfied easily.

Before moving on to computing the final baryon asymmetry, it is
worth mentioning that the $LH_u$ direction is free from the Q-ball
problem \cite{Fujii:2001dn}. Q-balls are non-topological solitons
which arise when the coherent oscillation of a flat-direction is
unstable against spatial perturbations. This typically happens
when the potential for the flat-direction is flatter than the
quadratic potential \cite{Kusenko:1997si}. If Q-balls are formed,
then all charges carried by the flat-direction are absorbed by the
Q-ball, hence the baryon asymmetry must be provided by the decay
of Q-balls, a situation typically disfavored for various reasons.
The supersymmetry breaking mass of the $LH_u$ direction, however,
has a big contribution from the large top yukawa coupling making
the potential for $\phi$ steeper than the quadratic potential,
thus avoiding the formation of Q-balls.

\subsection{The Baryon Asymmetry}\label{result}

We have seen in the previous subsection that a non-zero lepton
number is created during the inflaton-oscillation dominated epoch.
Since the $LH_u$ flat-direction also has a non-zero $B-L$ charge,
a non-zero $B-L$ number is also generated. An ${\cal O}(1)$
fraction of the $B-L$ number generated above is converted by
sphaleron effects to a non-zero baryon number since the $B-L$
number is generated at temperatures much above the electroweak
phase transition \cite{Kuzmin:1985mm}. More precisely, for a model with extra higgs
particles, one has:
\ba N_B = \frac{24+4N_H}{66+13N_H}\,(N_{B-L})\ea where $N_H$ is
the number of higgs doublets. Here, it has been implicitly assumed that all other new particles 
are much heavier. For an MSSM-like model with two
higgs doublets and in which the sparticles are camparable to the higgs masses,
the general expression has been computed in \cite{Chung:2008gv}. For our purposes, it will
suffice that an $\mathcal{O}(1)$ fraction of $B-L$ number is converted to $B$ number. 
As long as the condensate decays through $B-L$ conserving interactions after the
time it starts oscillating, this baryon number is intact and is
insensitive to the details of the decay. This will be assumed to
be the case. The baryon asymmetry $(\frac{n_B}{s})$ computed from
above is, however, diluted by other sources of entropy production
following inflation. The most natural source of such late-time
entropy production is the decay of moduli, alluded to in the
Introduction.

As was explained in section \ref{modulispectra}, string
compactifications which stabilize the Hierarchy as well as the
moduli at the same time generically contain a lightest modulus
(moduli) of ${\cal O}(m_{3/2})$. It is quite natural for the
lightest moduli $m_{X_{lightest}}$ to be displaced from their
minima (see section \ref{duringinf}). These moduli start
oscillating when the hubble parameter drops to $H \sim
m_{X_{light}}$ \footnote{We are ignoring thermal corrections since
these are typically quite small for moduli.}. These moduli evolve
as pressureless matter and hence their contribution relative to
the background radiation (from reheating after inflation) grows
rapidly. Hence, they quickly dominate the energy density of the
Universe. Also, since these moduli couple to the visible sector by
only gravitational (planck suppressed) interactions, they decay
quite late, very close to the beginning of BBN. Moduli such as
$Y_i$\footnote{these correspond to the throat moduli in Type IIB
compactifications.} which may be stabilized at an intermediate
scale ($H_I > M_{Y_i} > M_{X_{lightest}}$) do not affect the
result for the final baryon asymmetry since that depends only on
the entropy production from moduli which decay last, irrespective
of which moduli decay earlier.

The decay of lightest moduli long after inflation produces a lot
of entropy and dilutes the baryon asymmetry. Therefore, the final
baryon asymmetry can be estimated as: \ba \frac{n_B}{s}(final)
&\approx& \frac{N_B
R^{-3}}{\left(\rho^{X_{lightest}}/T_R^{X_{lightest}}\right)}\nonumber\\
&\approx& \frac{N_B\,H^2\,T_R^{X_{lightest}}}{3H^2\, m_p^2}\ea
Using (\ref{totalnL}), we find for the final baryon asymmetry:\ba
\frac{n_B}{s}(final) = {\cal
O}(1)\,\frac{M\,T_R^{X_{lightest}}}{m_p^2}\ea where we have used
conservatively that $\log(H_0/H_{osc})$ is ${\cal O}(1)$ in
(\ref{totalnL}) and that the second term in (\ref{totalnL}) is
suppressed. This is consistent with all the constraints in section
\ref{allconstraints}. We are only concerned with the estimate of
the baryon asymmetry, and the above result is true up to factors
of ${\cal O}(1)$. As argued in section \ref{modulispectra},
$m_{X_{light}} \gtrsim m_{3/2}$. In order to produce the desired
baryon asymmetry, one
requires: \ba \frac{n_B}{s}(final) &\sim& 10^{-10}\nonumber\\
\implies M \equiv \frac{\lambda\,M_s}{\hat{h}^2} &\sim&
10^{29}\;{\rm GeV} \ea Since $M_s\sim v_{B-L}\gtrsim 10^{16}$ GeV
for string constructions consistent with standard gauge
unification at $M_{GUT} \sim 10^{16}$ GeV, this implies: \ba
\label{reqmt}\left(\frac{\hat{h}^2}{\lambda}\right) \sim
10^{-12}\;\;{\rm for}\;\; M_s \sim 10^{17}\;{\rm GeV} \ea This is
consistent with the analysis in section \ref{postinf} for the
computation of the lepton number since terms proportional to
$\frac{\hat{h}^2}{\lambda}$ were assumed to be suppressed in the
computation. Note that the requirement of a tiny
$\frac{\hat{h}^2}{\lambda}$ amounts to having a virtually massless
lightest left-handed neutrino ($\sim 10^{-16}$ eV). We will argue
in section \ref{micro} that the above requirement for a tiny
$\frac{\hat{h}^2}{\lambda}$ (and hence a virtually massless
lightest left-handed neutrino) is quite natural to obtain in
string constructions in which string instantons generate the
required couplings. However, before doing that it is important to
address issues relating to the moduli(gravitino) problem and the
origin and abundance of dark matter within this framework.

\subsection{Moduli (Gravitino) Problem and Non-thermal Dark
Matter}\label{moduliDMproblems}

We first address the moduli and gravitino problems within this
framework. It is known that if the displaced light moduli have
masses $\gtrsim$ 10 TeV, then their decay reheats the Universe to
temperatures above a few MeV (assuming planck suppressed
interactions), allowing BBN to occur successfully. Within gravity
mediation, the mass of the lightest modulus (moduli) is comparable
to that of the gravitino as explained in section
\ref{modulispectra}. The precise spectrum depends on model-dependent details. 
One has to be careful, however, about the overproduction of gravitinos from the decay 
of the lightest modulus, if kinematically allowed. This could create problems for BBN as well as give rise to too much dark matter from their decays \cite{Nakamura:2006uc}. 
However, this problem can be elegantly solved if the lightest modulus (scalar field) $X_{0}$ has mass $m_{X_0} \approx m_{3/2} \gtrsim 10$ TeV. As explained in section \ref{modulispectra}, 
such fields are naturally available since there exist moduli which are not stabilized solely by non-perturbative effects.  In addition, constructions in which supersymmetry breaking is triggered by matter scalar fields in a dynamical supersymmetry breaking (DSB)
sector, also give rise to $m_{X_0}\approx m_{3/2}$
\cite{Lebedev:2006qq}. Therefore, the gravitino problem is
naturally avoided as the decay of $X_0$ to the gravitino is not
allowed kinematically. A similar mechanism is operative in the $M$
theory framework studied in \cite{Acharya:2008bk}. $X_0$ 
decays last and the baryon asymmetry is determined by the
reheat temperature of $X_0$. As mentioned above, as long as
$m_{X_0}\approx m_{3/2} \gtrsim 10$ TeV, $T_R^{X_0}$ is bigger
than a few MeV allowing BBN to proceed in the usual manner.

Within gauge mediation (presumably embedded in a string
compactification), there will again be (geometric) moduli coming
from the compactification; however their spectra depends on the
concrete embedding and is largely decoupled from low energy
phenomenology. In this situation, it is natural to expect that the
geometric moduli are reasonably heavier than $\sim 10$ TeV
implying that they decay well before BBN. However, there is still
the scalar partner of the goldstino (which we denote by $S$) whose
$F$-term breaks supersymmetry. In simple and generic models of
gauge mediation, the scalar $S$ is a little heavier
than $m_{3/2}$ given by $m_S \sim m_{3/2}\,(\frac{m_p}{\Lambda})$,
where $\Lambda$ is a high scale present in the k\"{a}hler
potential (see appendix \ref{gaugemodulus} for details). Since
this scalar has much stronger couplings to the visible sector
compared to that of geometric moduli, even for $m_{3/2} \lesssim
1$ GeV the scalar $S$ could decay before BBN and reheat the
Universe to temperatures above a few MeV \cite{Ibe:2007km}.

What can be said about the superpartner spectrum and dark matter
within this framework? Within gravity mediation, it has been shown
that if the modulus which couples to the gauginos is stabilized
from non-perturbative effects, then the mass of the gauginos is
suppressed relative to $m_{3/2}$ \cite{Conlon:2006us}. If in
addition, the supersymmetry breaking sector is sequestered from
the visible sector, then the scalars are also suppressed relative
to $m_{3/2}$, otherwise not. Therefore, the precise superpartner
spectrum will depend on these model-dependent details. It is
important to note, however, that even with $m_{3/2} \gtrsim 10$
TeV, it is naturally possible to have superpartners, particularly
gauginos, in the sub-TeV range. Within gauge mediation, the
gravitino is the LSP ($m_{3/2} \lesssim$ GeV) and hence the DM
candidate. The masses of other superpartners depend on the precise
mass scale of the messengers, but can easily take values in the
sub-TeV range.

Finally, we address the issue of the abundance of DM within string
compactifications in general and within the framework considered,
in particular. As explained in section \ref{modulispectra}, stabilizing
the moduli and generating a stable hierarchy between the
electroweak and planck scales at the same time requires that at
least some of the moduli are light, close to the TeV scale. Since the Hubble parameter 
during inflation is typically much larger than the TeV scale, the light moduli are generically displaced from their minima
during inflation and start oscillating when the hubble parameter
drops down such that $H \sim M_{moduli}$. These quickly dominate
the energy density of the Universe. Therefore, within string
compactifications which stabilize moduli and explain the
Hierarchy, the ``standard" thermal computation of relic dark
matter abundance has to be modified to take the effects of the
moduli into account. Since the couplings of visible sector fields
are determined by (at least some of) the light moduli, the moduli
couple to the visible sector fields and can decay into them.

Within gravity mediation, this means that the light moduli
decaying into superpartners will eventually decay to the LSP (DM
candidate). The most important such contribution will come from
the lightest moduli $X_0$. Since $m_{X_0} \gtrsim 10$ TeV, the
reheat temperature from decay of $X_0$ is $T_R^{X_0} \sim
\sqrt{\Gamma_{X_0}m_p} \sim {\cal O}(1)$ MeV, where $\Gamma_{X_0}
\sim \frac{m_{X_0}^3}{m_p^2}$ is the decay width of $X_0$. Since
the thermal freeze-out temperature of DM particles $T_f^{\chi_1}
\sim \frac{m_{\chi}}{25} \sim {\cal O}(1)$ GeV is typically much
larger than $T_R^{X_0}$, this  means that the DM particles
produced from decay of $X_0$ never reach thermal equilibrium,
thereby giving rise to a \emph{non-thermal} abundance of dark
matter. The thermal abundance of dark matter from the thermal
plasma after inflation is greatly diluted by entropy production
from decay of the various moduli. Therefore, the non-thermal
abundance generically dominates over the thermal one. Scenarios of 
non-thermal Dark Matter which fit within this framework have been explored in 
\cite{Acharya:2008bk, Endo:2006ix}.

Within gauge mediation, the situation is quite similar except that
the gravitino is the DM candidate and is stable. The decay of
geometric moduli before the decay of $S$ will again greatly dilute
the thermal abundance of the gravitinos. The scalar $S$ can
however, decay directly into gravitinos giving rise to a
non-thermal abundance of DM\footnote{If staus are NLSPs, they can
also be produced from the decay of $S$ if kinematically allowed.
BBN constraints then typically imply an upper bound for the
gravitino mass \cite{Ibe:2007km}. The precise value is
model-dependent.}, similar to that in gravity mediation. This is
due to the fact that the reheat temperature resulting from the
decay of $S$ is $T_R^{S}\gtrsim{\cal O}(10)$ MeV for ``reasonable"
values of $m_{3/2}$ and $\Lambda$ \footnote{The precise number is
model-dependent and depends on details. See appendix
\ref{gaugemodulus} for more discussion.}, which is much smaller
than the typical freeze-out temperature $T_f^{(3/2)}$ of
gravitinos implying that the gravitinos produced never reach
thermal equilibrium.

Depending on the details of the particular construction, the yield
of DM particles from the decay of $X_{lightest}$ could be above or
below that of the so-called ``critical" density at
$T=T_R^{X_{lightest}}$: \ba n_{\chi}^{(c)} = \frac{3H}{\langle
\sigma_{\chi} v\rangle}|_{T_R^{X_{lightest}}} \ea If the yield is
above the critical yield, the DM particles quickly annihilate
until they reach the critical density above. On the other hand, if
the yield of DM particles is below the critical yield, then the
comoving abundance ($Y_{\chi} \equiv n_{\chi}/s$) is given by: \ba
Y_{\chi} \sim
\frac{B^{\chi}_{X_{lightest}}n_{\chi}^{(0)}}{(T_R^{X_{lightest}})^3}\ea

In both situations, and both within gravity and gauge mediation,
the final relic density depends on the underlying physics
determining the coupling of the lightest modulus to visible sector
particles. Therefore, the upper bound on the observed relic
density serves as an important constraint on the moduli-matter and
moduli-gravitino coupling in string constructions. It has been
argued that such non-thermal production mechanisms can provide the
correct DM abundance in some string frameworks, such as the one in
\cite{Acharya:2008bk}.

\section{Microscopics}\label{micro}

\subsection{$U(1)_s$, Anomalies, Gauge Invariance and Brane
Instantons}\label{summaryanomaliesU1}

In this section, we give a brief account of Type II string
constructions in which perturbatively absent couplings (forbidden
by gauge and global symmetries) can be generated by stringy
effects and which have been studied in the literature. Examples of
such couplings include yukawa couplings, the $\mu$ parameter,
majorana mass terms for neutrinos and certain non-renormalizable
operators \cite{Blumenhagen:2006xt}. In this work, we will
primarily be interested in right-handed neutrino majorana masses
and the $B-L$ violating ``Weinberg" operator leading to neutrino
masses.

Within the setup of Type IIB orientifold flux compactifications
with D3/D7 branes and orientifold planes (in which moduli
stabilization is best understood), semi-realistic matter spectra
can be constructed with D3 branes at singularities and/or with D7
branes wrapping supersymmetric four-cycles. Important consistency
constraints arise from the requirement of tadpole
cancellation\footnote{this is bascially a generalization of
Gauss's law for fluxes.}. In particular, in order to have
\emph{chiral} spectra from D7-branes, the $D7$-branes must be
equipped with a non-trivial magnetic flux on their world-volume.
It is known that a chiral spectrum is anomalous in general.
However, it can be shown that the requirement of tadpole
cancellation actually guarantees the absence of non-abelian gauge
anomalies. The cancellation of pure abelian and mixed
abelian-non-abelian(graviton) anomalies is more subtle. The
cancellation of these potential anomalies is guaranteed in string
theory by the Green-Schwarz mechanism, as follows.

The action for a D7-brane (D7$_a$) contains a Chern-Simons term
(among others) of the form: \ba S_{CS} \sim
\int_{\mathbf{R}^{3,1}\times \Sigma_a}\,C_4\wedge {\cal F} \wedge
{\cal F} \ea where $C_4$ is the RR 4-form in Type IIB string
theory and ${\cal F}$ is a two-form. Taking one of the ${\cal F}$
along $\Sigma_a$ (part of world-volume flux on D7$_a$) and the
other ${\cal F}$ to be the field strength of the $U(1)_a$ gauge
field on D7$_a$, and expanding $C_4$ as: \ba C_4 = C_2^{\alpha}
\wedge \omega_{\alpha}+...\ea gives under certain topological
conditions: \ba S_{CS} \sim
\int_{\Sigma_a}\,\omega_{\alpha}\wedge{\cal
F}\int_{\mathbf{R}^{1,3}}\,C_2^{\alpha}\wedge F_a \sim
\alpha'^2\,Q_{\alpha}^a\,\int_{\mathbf{R}^{1,3}}\,C_2^{\alpha}\wedge
F_a\ea with $Q_{\alpha}^a$ an integer-valued topological charge
matrix. The $C_2^{\alpha}\wedge F_a$ coupling leads to two effects
- a) it provides a Stuckelberg mass term for the $U(1)_a$ gauge
field, and b) the shift symmetries of the axions
($a^{\alpha}\equiv {\rm Im}(T^{\alpha})$) which are dual to the RR
2-forms $C_2^{\alpha}$ are gauged. The gauged shift symmetry of
the axions gives rise to a $D$-term for the corresponding $U(1)$s
as in (\ref{Dterm}).

The $U(1)_a$ gauge fields which remain massless lie in the
kernel\footnote{The kernel of a matrix ${\bf Q}$ is the set of all
eignevectors ${\bf x}$ for which ${\bf Q\,x}=0$.} of the matrix
$Q_a^{\alpha}$. These $U(1)_s$ are always anomaly free. For
example, in realistic string constructions $U(1)_Y$ must always be
massless; so it must lie in the kernel. On the other hand, massive
$U(1)_a$ gauge fields could be either anomalous or anomaly free
depending on the details. For anomalous $U(1)$s, there appears an
additional term in the effective action -
$\int_{\mathbf{R}^{1,3}}\,a^{\alpha}\,{\rm tr}(F^{\alpha}\wedge
F^{\alpha})$ which precisely cancels the mixed abelian-non-abelian
gauge(graviton) anomalies alluded to above. The above coupling is
not present for non-anomalous $U(1)$s since these are already
anomaly free by definition. Non-anomalous massive $U(1)$s are
quite interesting as they provide a way to reduce the gauge
symmetry at low energies without the Higgs mechanism. A natural
example of such a massive anomaly-free $U(1)$ is $U(1)_{B-L}$.
This can naturally generate majorana neutrino masses and the $B-L$
violating weinberg operator, as we will see below.

Focussing on string constructions with a $U(1)_{B-L}$ gauge
symmetry, we are interested in the case when the details of the
compactification lead to a structure where the two forms
$C_2^{\alpha}$ (or their duals $a^{\alpha}$) couple to the
$U(1)_{B-L}$ gauge field in such a way as to provide a Stuckelberg
mass term for the gauge field. The massive $U(1)_{B-L}$ still
survives at low energies as a perturbative global symmetry.
Therefore, majorana neutrino masses, among other operators, are
forbidden at the perturbative level. However, non-perturbative
effects may exist which violate this global symmetry to a discrete
subgroup\footnote{In many cases, it turns out that the discrete
subgroup is none other than the usual $R$-parity
\cite{Ibanez:2007rs}.} and generate these operators. Such a
non-perturbative effect actually does exist in string theory; it
is provided by (euclidean) brane instantons. Because the effective
perturbative $U(1)_{B-L}$ global symmetry arising at low energies
is secretly gauged in string theory, the operator induced by the
brane instantons must respect the underlying gauge invariance.
This implies, in particular, that the non-perturbatively generated
majorana mass parameter ($M_N$) should transform under
$U(1)_{B-L}$ in such a way as to make the operator $M_N\,NN$
$U(1)_{B-L}$ gauge invariant. This mechanism is quite general and
can formally occur in Type IIA, Type IIB, Heterotic and $M$ theory
constructions. Most of the work related to model-building in this
regard is done in Type IIA constructions \cite{Ibanez:2007rs}
\footnote{see \cite{Ibanez:2007tu} for some work in local Type IIB
constructions.} where it has been shown that there exists a large
set of models satisfying the various criteria required for the
above mechanism to work. Since Type IIA constructions are related
by T-duality to Type IIB constructions, we expect a large set of
models incorporating the above mechanism to exist within Type IIB
constructions as well. Assuming this to be the case, we will keep
working within the framework of Type IIB compactifications with
D3/D7 branes\footnote{This is because moduli stabilization is best
understood in this setup.}.

We briefly describe the microscopic mechanism by which the brane
instanton generates the relevant operators and the conditions
which need to be satisfied. In computing the spacetime interaction
mediated by the instanton, one has to integrate over the instanton
zero modes. The brane instanton relevant in Type IIB
compactificatiosn are E3 instantons wrapping supersymmetric
four-cycles in the compact space and invariant under the
orientifold projection. In order to contribute to the effective
superpotential, the instanton must have the right structure of
zero modes. Generically, one only has universal bosonic zero
modes, those corresponding to the position of the instanton. The
fermionic zero modes are much more important though. The uncharged
(under the 4D gauge group) fermionic zero modes have to satisfy
certain conditions to contribute to the superpotential
\cite{Blumenhagen:2006xt}. We will focus on (chiral) charged
fermionic zero modes which are much more interesting and relevant.
Microscopically, these charged fermionic zero modes arise from
open strings at the intersection of the E3 instanton $E$ with
spacetime-filling magnetized D7-branes $D7_i$ (and their
orientifold images $D7_i'$) present in the construction. The net
number of these zero modes is given by $I_{Ei}-I_{Ei'}$, where
$I_{Ei}$ is the intersection number between $E$ and $D7_i$. In
order to saturate the integration over the charged fermionic zero
modes, the spacetime interaction must therefore contain insertions
of charged (in particular under $U(1)_s$) 4D fields, giving rise
to a superpotential of the following general form:
\ba\label{matterE3} W \sim e^{-S_{E3}}\,\phi_1\,\phi_2...\phi_n\ea
The action of the instanton is proportional to the volume of the
four-cycle which it wraps. This is measured by a k\"{a}hler
modulus ($T^{E}\equiv\tau^E+ia^E$), whose imaginary part is an
axion $a^{E}$ which shifts as described in the previous section
since the D7-branes are magnetized. \ba
{\rm Re(S_{E_3})} &\sim& \frac{2\pi}{g_s}\frac{{\rm Vol}(\Sigma_4)}{l_s^4} \equiv {2\pi}\,\tau^E\nonumber\\
{\rm Im(S_{E_3})} &\sim&{2\pi}\,a^E\ea From the structure of zero
modes and Stuckelberg couplings of the type
$\int\,C_2^{\alpha}\wedge F_a$, one can show that under $U(1)_X
=\sum_i\,U(1)_i$: \ba e^{-S_{E3}} &\rightarrow&
e^{-i\sum_i\,(I_{Ei}-I_{Ei'})\Lambda_i}\,e^{-S_{E3}} \nonumber\\
\phi_1\phi_2...\phi_n &\rightarrow&
e^{i\sum_i\,(I_{Ei}-I_{Ei'})\Lambda_i}\, \phi_1\phi_2...\phi_n \ea
Thus the $E3$-instanton transforms under $U(1)_X$ in such a way as
to precisely cancel the transformation of the charged matter
fields, making the superpotential in (\ref{matterE3}) $U(1)_X$
gauge invariant. From the low energy effective field theory point
of view however, the $U(1)_X$ appears as a global symmetry which
is broken by the $E3$ instanton\footnote{generally to a discrete
subgroup.}.

\subsection{Application to Neutrinos}\label{applyneutrinos}

We can now apply the above formalism to the special case of
majorana neutrino masses and the weinberg operator with
$U(1)_X=U(1)_{B-L}$. This has been done in the literature. The
zero mode structure required to obtain the two operators is
different for the two cases \cite{Ibanez:2007rs}. Hence, two
different kinds of $E3$ instantons contribute to the two
operators. Depending on the details of the construction, one or
both kinds of instantons may contribute to neutrino masses. We
will look at both of them in detail.

When the structure of zero modes satisfies a certain condition
\cite{Ibanez:2007rs},
one gets the following operator: \ba \label{WM}W_{M} &=& M_N^{ab}\,N_R^a N_R^b \nonumber\\
{\rm where}\;\,M_N^{ab} &=& m_p\sum_r\,d_r^a
d_r^b\,e^{-S_{E3}^{(M),r}}\ea where $d_r^{a,b}$ are
model-dependent coefficients of ${\cal O}(1)$. In general, many
instantons (with the same zero mode structure) contribute to the
majorana mass parameter. The summation in (\ref{WM}) illustrates
that fact. The flavor structure of $(\ref{WM})$ is such that
having three or more instantons gives rise to three non-zero
eigenvalues. Since the volumes wrapped by these instantons will
generically differ by ${\cal O}(1)$, the three mass eigenvalues of
the matrix $M_N^{ab}$ will generically be hierarchical, as stated
in section \ref{duringinf}\cite{Ibanez:2007tu}. This will thus
lead to the following left-handed neutrino masses:
\ba\label{seesaw} M_L^{ab}({\rm seesaw})= \langle
H_u^0\rangle^2\,(\mathbf{h^T\,M_N^{-1}h})^{ab}\ea

The Weinberg operator, which requires a different structure of
zero modes \cite{Ibanez:2007rs}, is given by: \ba\label{wein} W_W
&=& \kappa\,\frac{LH_uLH_u}{m_p}\nonumber\\\implies M_L^{ab}({\rm
weinberg}) &=& \frac{\langle H_u^0\rangle^2}{m_p}\sum_r c^a_r
c^b_r e^{-S_{E3}^{(W),r}}\ea Again, many different instantons
(with the same zero mode structure) could contribute to the
left-handed neutrino masses and naturally lead to small
left-handed neutrino masses.

As mentioned above, depending on the details one or both kinds of
instantons may contribute to the neutrino masses. It could also
happen that one may dominate the other. Note however, that in
string vacua with $N_R$'s massless at the perturbative level, one
definitely requires instantons to make the $N_R$'s massive at a
phenomenologically acceptable level. Therefore, here we will
assume that both kinds of $E3$ instantons discussed above exist.
In the explicit string constructions considered in
\cite{Ibanez:2007rs}, such examples were found in large
number\footnote{Large number of models were found after satsifying
a relaxed constraint on the symmetries of the instantons which can
easily occur in the presence of fluxes \cite{Ibanez:2007rs}.}. We
will consider both cases where the instanton generating majorana
masses dominates over the one generating the weinberg operator and
vice versa, and argue that it is naturally possible to obtain the
required value of $(\frac{\hat{h}^2}{\lambda})$ in both cases.
Also, since we are primarily interested in the baryon asymmetry,
we will mostly focus on the {\it lightest} left-handed neutrino
mass. In section \ref{possibleLHCconnection}, however, a very
interesting possibility will be briefly described in which one has
${\cal O}$(TeV) right handed neutrinos consistent with high-scale
affleck-dine leptogenesis and the moduli problem. Potential
consequences of these for Dark Matter and the LHC will also be
briefly discussed.

Before moving on to the details, we would like to comment on the
flavor structure of the neutrino mass matrix. The detailed flavor
structure is model-dependent; however some general aspects of the
flavor structure within this context have been studied in
\cite{Antusch:2007jd} where it was shown that it is possible to
incorporate a flavor structure for the neutrino mass matrix
consistent with observations with reasonable assumptions. Since we
are not interested in the detailed flavor structure for our
purposes, for concreteness and simplicity we will assume the
``normal hierarchy" case with sequential dominance
\cite{Antusch:2004gf}. This is also natural from a theoretical
point of view since the different instanton contributions are
naturally hierarchical as argued above. In fact, within the
paradigm of sequential dominance, it is quite natural to have only
two instantons dominantly give rise to the observed neutrino
masses and mixings making the lightest left-handed neutrino
virtually massless\cite{Antusch:2007jd,Antusch:2004gf}. This is
\emph{precisely} what is required for achieving a large enough
baryon asymmetry within the framework studied here. It has to be
kept in mind though that other flavor structures are also
possible. Those can also be probably incorporated within this
framework.

\subsection{Microscopic Constraints to obtain a small
$\frac{h^2}{\lambda}$}\label{constraints}

In this subsection, we will look at the computation of the
lightest neutrino mass in detail. The expression for the
left-handed neutrino mass matrix was given in (\ref{seesaw}) and
(\ref{wein}). However, those expressions are not canonically
normalized. One has to be careful in properly normalizing the
matter fields such as to give rise to a canonical kinetic term for
these fields. This is especially important within supergravity
since the K\"{a}hler potential for these fields is non-canonical
in general. In general, if the (un-normalized) superpotential is
given by: \ba W =
M_{\alpha\beta}\phi_{\alpha}\phi_{\beta}+Y_{\alpha\beta\gamma}
\phi_{\alpha}\phi_{\beta}\phi_{\gamma}+\frac{\kappa_{\alpha\beta\gamma\delta}}{m_p}
\phi_{\alpha}\phi_{\beta}\phi_{\gamma}\phi_{\delta}+...,\ea then
the canonically normalized quantities in $W$ above are
\cite{Brignole:1997dp}: \ba \label{normalized}
\hat{M}_{\alpha\beta} &=& e^{K/2}\,M_{\alpha\beta}(\tau_m)\nonumber\\
\hat{Y}_{\alpha\beta\gamma} &=&
e^{K/2}\,\frac{Y_{\alpha\beta\gamma}(U_m)}{(\tilde{K}_{\alpha}\tilde{K}_{\beta}\tilde{K}_{\gamma})^{1/2}(\tau_m,U_m)}
\approx
e^{K/2}\,\frac{Y_{\alpha\beta\gamma}(U_m)\,(g^{-1}_{\alpha}g^{-1}_{\beta}g^{-1}_{\gamma})(U_m)}
{(\tilde{K}^0_{\alpha}\tilde{K}^0_{\beta}
\tilde{K}^0_{\gamma})^{1/2}(\tau_m)}\;\;({\rm no\;sum})\nonumber\\
\hat{\kappa}_{\alpha\beta\gamma\delta}&=&
e^{K/2}\,\frac{\kappa_{\alpha\beta\gamma\delta}(U_m)}{(\tilde{K}_{\alpha}\tilde{K}_{\beta}
\tilde{K}_{\gamma}\tilde{K}_{\delta})^{1/2}(\tau_m,U_m)} \approx
e^{K/2}\,\frac{\kappa_{\alpha\beta\gamma}(U_m)\,(g^{-1}_{\alpha}g^{-1}_{\beta}g^{-1}_{\gamma}g^{-1}_{\delta})(U_m)}
{(\tilde{K}^0_{\alpha}\tilde{K}^0_{\beta}
\tilde{K}^0_{\gamma}\tilde{K}^0_{\delta})^{1/2}(\tau_m)}\;\;({\rm
no\;sum})\ea In the above, we have shown the moduli dependence of
the various parameters and have used the fact that the k\"{a}hler
metric $\tilde{K}_{\alpha}$ for matter fields $\phi_{\alpha}$ can
be factorized as
$\tilde{K}_{\alpha}(\tau_m,U_m)=\tilde{K}^0_{\alpha}(\tau_m)\,g_{\alpha}(U_m)$\footnote{For
simplicity, we have also assumed that the k\"{a}hler metric is
roughly diagonal, i.e. $\tilde{K}_{\bar{\alpha}\beta}\approx
\tilde{K}_{\alpha}\delta_{\alpha\beta}$. This will not change the
main conclusion in the analysis.} \cite{Conlon:2007dw}. Applying
the above formulas to the majorana mass operator and the weinberg
operator, one gets the following contributions to the left-handed
neutrino mass: \ba\label{seesaw-wein} \hat{M}^L_{(\nu_{ab})}({\rm
seesaw}) &=& {\cal O}(1)\,\frac{\langle
H_u^0\rangle^2}{m_p}\,\frac{e^{K/2}}{\tilde{K}^0_{L}\tilde{K}^0_{H_u}
\tilde{K}^0_{N_3}}\frac{({\bf h_D}^T\,{\bf h_D})}{(\sum_r d^r_ad^r_b\,e^{-S_{E3}^{(M),r}})}\nonumber\\
\hat{M}^L_{(\nu_{ab})}({\rm weinberg}) &=& {\cal
O}(1)\,\frac{\langle
H_u^0\rangle^2}{m_p}\,\frac{e^{K/2}}{\tilde{K}^0_{L}\tilde{K}^0_{H_u}}(\sum_r
c^r_a c^r_b\,e^{-S_{E3}^{(W),r}})\ea Here, we have used the fact
that generically $\tilde{K}_{\alpha}(\tau_m,U_m)={\cal
O}(1)\tilde{K}^0_{\alpha}(\tau_m)$ and have also used a diagonal
neutrino yukawa coupling for simplicity (${\bf h} \approx {\bf
h_D}$). One can further simplify the above expressions if one
assumes that the various instantons contributing to the operators
above are hierarchical, which we have argued to be quite natural.
Thus, the sum over the instantons in both operators in
(\ref{seesaw-wein}) above is dominated by only one instanton.
Using the fact that typically $c_a^r,d_a^r = {\cal O}(1)$, this
gives rise to the following expression for the lightest
left-handed neutrino mass: \ba\label{lightest}
\hat{M}^L_{(\nu_1)}({\rm seesaw}) &=& {\cal O}(1)\,\frac{\langle
H_u^0\rangle^2}{m_p}\,\frac{e^{K/2}}{\tilde{K}^0_{L}\tilde{K}^0_{H_u}
\tilde{K}^0_{N_3}}{({h_D^1})^2}{(e^{S_{E3}^{(M)}})}\nonumber\\
\hat{M}^L_{(\nu_1)}({\rm weinberg}) &=& {\cal O}(1)\,\frac{\langle
H_u^0\rangle^2}{m_p}\,\frac{e^{K/2}}{\tilde{K}^0_{L}\tilde{K}^0_{H_u}}(e^{-S_{E3}^{(W)}})\ea

For later purposes, it is useful to write down expressions for the
dependence of the string scale $M_s$, the k\"{a}hler potential and
the k\"{a}hler metric on the volume (k\"{a}hler moduli). These are
given by \cite{Conlon:2006tj}:
\ba\label{metric} M_s &=& \frac{m_p}{\mathcal{V}^{1/2}}\nonumber\\
K&=&-2\log{\mathcal{V}}+...\nonumber\\ \tilde{K}_{\phi} &=&
{\tau_m}^{-1/2}{\cal O}(1) = {\cal V}^{-1/3}{\cal O}(1)\ea where
for the k\"{a}hler metric we have assumed for simplicity that all
four-cycles in the Calabi-Yau are roughly of the same size. Since
we are interested in models with standard gauge unification, a
natural choice is $M_{GUT}\lesssim M_s \sim 10^{17}$GeV, implying
that ${\cal V}^{1/2}={\cal O}(10)$. Also, for the supergravity
approximation to hold, we will require that the moduli are
stabilized at values greater than unity, but still respecting the
constraint that ${\cal V}^{1/2}={\cal O}(10)$. We will use all
this in our subsequent analysis.

\subsubsection{Majorana Operator Domination (see-saw case)}\label{caseM}

We will first consider the case where the see-saw contribution
dominates over that of weinberg one. From (\ref{seesaw-wein}),
this requires: \ba\label{seesawbig} (h_D^1)^2 &\gg&
e^{-(S_{E3}^{(M)}+S_{E3}^{(W)})}\,\tilde{K}^0_{N_3}\ea In
addition, in order to obtain the correct baryon asymmetry
(\ref{reqmt}) in this case, one requires: \ba\label{seesawrqmt}
\left(\frac{\hat{h}^2}{\lambda}\right)_{{\rm seesaw}} =
\frac{(h_D^1)^2e^{K/2}}{(\tilde{K}^0_{L}\tilde{K}^0_{H_u}
\tilde{K}^0_{N_3})\,e^{-\frac{2\pi}{g_s}{\tau}^{N_3}}}\,\frac{M_s}{m_p}
\sim 10^{-12}\ea Here we have used the fact
$\hat{M}_{N_3}=e^{K/2}e^{-2\pi{\tau}_{N_3}}\,m_p =
\hat{\lambda}\,M_s$, implying
$\hat{\lambda}=\frac{e^{K/2}e^{-2\pi\tau_{N_3}}\,m_p}{M_s}$. Using
(\ref{metric}), one gets: \ba
\label{yukawa}(h_D^1)^2e^{2\pi\tau_{N_3}}\sim 10^{-12}{\mathcal
V}^{1/2}{\cal O}(1) \ea The value of $\tau_{N_3}$ is constrained
by the requirement that the right-handed sneutrino
$\hat{\tilde{N}}_3$ is not displaced from its original minimum in
order to produce a non-zero lepton number as argued in appendix
\ref{fateofN3}. Conservatively, this means that $M_{N_3}$ is
larger than the hubble parameter during inflation, i.e.\ba
\label{MN3}\hat{M}_{\tilde{N}_3} &\approx& M_{N_3} = {\cal O}(1)\,
e^{K/2}\,e^{-{2\pi}{\tau_{N_3}}}\,m_p \gtrsim H_I\nonumber\\
\implies \frac{e^{-{2\pi}{\tau_{N_3}}}}{\mathcal{V}}\,m_p
&\gtrsim& H_I \approx 10^{12}-10^{13}\,{\rm GeV}\ea This implies
that $\tau_{N_3}$ be close to (but greater than) unity for ${\cal
V}^{1/2}={\cal O}(10)$. This is still consistent with the
supergravity approximation.

The un-normalized yukawa couplings $h_D$ arise from multiply
wrapped world-sheet instantons and are given by products of Jacobi
theta functions \cite{Cremades:2004wa}. These can be well
approximated as $h_D^{ijk} \approx {\cal O}(1)e^{-\eta^{ijk}4\pi
U_m}$ where $\eta^{ijk}={\cal O}(1)$. These yukawas depend on the
complex structure moduli ($U_m$) in Type IIB string theory which
are stabilized by bulk fluxes and can take a wide range of values
depending on the fluxes. From (\ref{yukawa}) and (\ref{MN3}), one
finds that $(h_D^1)^2 \gtrsim 10^{-15}$. This can be naturally
obtained by $U_m ={\cal O}(1)$(but greater than
unity)\footnote{The {\cal O}(1) factors at various places allow
multiple ways of satisfying the constraint.}.

Also, the condition (\ref{seesawbig}) that the majorana operator
dominates the weinberg operator gives rise to a constraint on
$\tau_W$. Using (\ref{yukawa}), the condition (\ref{seesawbig})
gives: \ba
e^{-2\pi \tau_W} &\ll& 10^{-10}\nonumber\\
\implies &\tau_W \gtrsim& 3.6\ea which can again be satisfied
naturally. Finally, we have to check that for the above choice of
parameters the quantity $(\frac{\hat{h}}{\lambda})$ is suppressed
as has been assumed in the analysis of section \ref{postinf}. One
finds that: \ba\left(\frac{\hat{h}}{\lambda}\right)_{\rm
see\,saw}= \frac{(h_D^1)}{(\tilde{K}^0_{L}\tilde{K}^0_{H_u}
\tilde{K}^0_{N_3})^{1/2}\,e^{-{2\pi}{\tau}^{N_3}}}\,\frac{M_s}{m_p}
\lesssim 10^{-3}\ea which is adequately suppressed.

\subsubsection{Weinberg Operator Domination}\label{caseW}

We now turn to the other case. This gives: \ba \label{weinbig}
(h_D^1)^2\, &\ll&
e^{-(S_{E3}^{(M)}+S_{E3}^{(W)})}\,\tilde{K}_{N_3}\ea In order to
obtain the correct baryon asymmetry in this case, one requires:
\ba \label{constrW} \left(\frac{\hat{h}^2}{\lambda}\right)_{{\rm
weinberg}} &=&
\frac{e^{K/2}}{\tilde{K}_{L}\tilde{K}_{H_u}}\,e^{-{2\pi}\tau_W}\,\frac{M_s}{m_p}
\sim 10^{-12}\nonumber\\{\rm or}\;
\frac{e^{-{2\pi}\tau_W}}{\mathcal{V}^{5/6}}&\sim&
10^{-12};\;\;\;{\rm using}\;(\ref{metric})\nonumber\\
\implies \tau_W &\approx& 3.5-4 \ea The requirement for the
majorana mass of $\tilde{N}_3$ is the same as in the previous
case: \ba\label{N32} \hat{M}_{\tilde{N}_3} \approx \hat{M}_{N_3}
&=& {\cal O}(1)\, e^{K/2}\,e^{-{2\pi}{\tau_{N_3}}}\,m_p \gtrsim
H_I\nonumber\\ e^{-2\pi\tau_{N_3}} &\gtrsim& 10^{-4}\ea again
leading to $\tau_{N_3}$ close to (but greater than) unity. Using
(\ref{constrW}) and (\ref{N32}), the condition (\ref{weinbig}) for
weinberg operator domination gives: \ba\label{yuk2} (h_D^1)^2
&\ll& \frac{10^{-16}}{{\cal V}^{1/2}} \approx 10^{-17}\;\;{\rm
using}\;(\ref{metric})\ea  So, using $h_D^{ijk} \approx {\cal
O}(1)e^{-\eta^{ijk}4\pi U_m}$ where $\eta^{ijk}={\cal O}(1)$, one
finds that (\ref{yuk2}) requires $U_m \gtrsim 1.5$. Finally, one
can estimate: \ba \left(\frac{\hat{h}}{\lambda}\right)_{\rm
weinberg} \sim \frac{(h_D^1)}{(\tilde{K}^0_{L}\tilde{K}^0_{H_u}
\tilde{K}^0_{N_3})^{1/2}\,e^{-{2\pi}{\tau}_{N_3}}}\,\frac{M_s}{m_p}
\ll 10^{-4}\ea which is consistent with our assumption.

Thus, we find that the range of the values of the stabilized
moduli in both cases are different from each other. Nevertheless,
both sets of ranges are perfectly natural to obtain from
perturbative string compactifications which are consistent with
the supergravity approximation and a compactification scale close
to the unification scale $M_G$. Finally, it is important to
remember that it could happen that both contributions to neutrino
masses are comparable to each other. A much bigger parameter space
for the stabilized values of the moduli opens up in the general
case, although the analysis is more complicated. Qualitatively
however, the analyses in the previous subsections show that it is
possible to obtain the required baryon asymmetry with moduli fixed
in the supergravity regime and a small neutrino yukawa coupling.

\section{Other String/$M$ Theory Compactifications}\label{other}

In this work, we have mostly focussed on Type IIB
compactifications. However, as long as the relevant microscopic
criteria are satisfied, the mechanism described in this paper
could work in other kinds of string/$M$ theory compactifications
which stabilize the moduli and generate the Hierarchy at the same
time. One such example in which it is naturally possible to
achieve the above is $M$ theory compactifications studied in
\cite{Acharya:2006ia}. Examples in other corners of string/$M$
theory may exist as well. Within $M$ theory compactifications, it
was shown in \cite{Acharya:2006ia} that with some reasonable
assumptions, strong gauge dynamics in the hidden sector can
stabilize all the bulk moduli as well as generate the Hierarchy in
a natural manner in a de Sitter (dS) vacuum. Supersymmetry
breaking is mediated to the visible sector by gravity. The moduli
spectra can then be computed reliably and it turns out that the
lightest moduli have masses $M_{X_{lightest}}\approx m_{3/2}$
which decay close to (but earlier than) BBN generating a large
amount of entropy \cite{Acharya:2008bk}. Therefore, the
microscopic criteria required in order to produce the correct
baryon asymmetry by the Affleck-Dine mechanism in this case are
precisely the same as in the Type II case. Also, since $m_{3/2}$
can naturally be greater than $\sim 10$ TeV and supersymmetry
breaking is mediated by gravity, there is no moduli (gravitino)
problem within this framework. It is known that $M$ theory
compactified on a singular seven dimensional space with $G_2$
holonomy is dual, in different regions of its moduli space, to
both heterotic string theory on an appropriate Calabi-Yau
\cite{Acharya:2001gy} and Type IIA string theory on another
appropriate Calabi-Yau with D6-branes and O6-planes
\cite{Atiyah:2001qf}. Since many examples of heterotic and Type
IIA string models with an $SM \times U(1)_{B-L}$ gauge group and
an MSSM-like chiral spectrum (with right-handed neutrinos) exist
in literature, this means that in principle this is possible in
the $M$ theory constructions as well. Finally, masses, yukawa
couplings and non-renormalizable terms like the Weinberg operator
in $M$ theory arise from membrane instantons connecting the
various superfields \cite{Acharya:2001gy, Atiyah:2001qf}. So, the
appropriate zero-mode structure has to be satisfied in order for
the instantons to contribute to the relevant terms in the
superpotential, similar to that in Type II constructions. It would
be extremely interesting to have explicit constructions satisfying
the above criteria.

\section{Potential Consequences for Observable Physics
} \label{possibleLHCconnection}

It would be very interesting to look for possible signals in particle physics and cosmology 
experiments in order to test the ideas proposed in this paper. Although precise details 
are model-dependent, one can still make general observations which are nevertheless quite interesting. As already pointed out in section \ref{moduliDMproblems}, the framework generically
predicts a non-thermal mechanism for the production of Dark Matter. One consequence of this is 
that DM candidates with a much larger annihilation cross-section\footnote{by a factor of upto $\mathcal{O}(1000)$.}, compared to that required for a standard thermal relic abundance, are allowed \cite{Acharya:2008bk}. This is essentially because the final relic abundance is set by the physics of the late-decaying scalar. This result is quite intriguing in the sense that a DM interpretation of recent results from DM indirect detection experiments like PAMELA and ATIC also require much
larger annihilation cross-sections. If the DM interpretation survives, it might be taken as a possible hint for non-thermal Dark Matter\footnote{Other explanations are also possible.}. 

Within this framework, a very interesting possibility could exist 
in which one could have ${\cal O}$(TeV) right-handed neutrinos
compatible with the production of baryon number at times much
earlier than that of the electroweak phase transition. Moreover,
the ${\cal O}$(TeV) right-handed neutrino will not be accompanied
by a low energy $U(1)_{B-L}$. Although such a light right-handed
neutrino is not a definite prediction of the framework, it is
quite interesting and hence worth some attention because it has
consequences for Dark Matter as well as the LHC.

As already explained, the right-handed neutrino masses are
naturally hierarchical since they are suppressed exponentially
relative to one another. Therefore, it is quite possible to have
the lightest right-handed neutrino to be light, near the
electroweak scale. Depending upon which operator (majorana or
weinberg) is dominantly responsible for the heaviest left-handed
neutrino mass (for which there is an experimental upper bound),
different parameter spaces are available for the stabilized values
of the moduli such that the lightest right-handed neutrino is
around the electroweak scale\footnote{remember we have assumed
sequential dominance of right-handed neutrinos.}. If both
operators are comparable, a much bigger parameter space opens up.
The requirement that the lightest right-handed neutrino be around
the electroweak scale implies: \ba \hat{M}_{N_1} = {\cal
O}(1)\,\frac{e^{-{2\pi}\tau_{N_1}}\,m_p}{\mathcal{V}}
&\sim& 10^2\;{\rm GeV}\nonumber\\
\implies \tau_{N_1} &\approx& 5.1\ea Note that the above could be
true irrespective of which operator, majorana or weinberg, is
responsible for the lightest left-handed neutrino mass $M_{\nu_1}$
and hence the baryon asymmetry $n_B/s$. Thus, it is
possible\footnote{note that this is a natural possibility but not
a concrete prediction.} for the baryon asymmetry to be produced
much before the electroweak phase transition while still having an
electroweak scale right-handed neutrino.

If such a light right-handed neutrino exists, it has interesting
consequences for the LHC as well as Dark Matter. To understand
this better, it is useful to write the relevant interactions.
Assuming that the neutrino yukawas are roughly diagonal as before,
we can write the relevant interactions containing the lightest
right-handed neutrino as: \ba \label{lightrhneutrino}W &\supset&
\hat{h}_{3}\hat{N}_1\hat{L}_3\hat{H}_u +
\hat{M}_{N_1}\,\hat{N}_1\hat{N}_1 \nonumber\\
\mathcal{L}_{soft} &\supset&
\tilde{m}^2\hat{\tilde{N}}_1^{\dag}\hat{\tilde{N}}_1+
(\hat{A}_3\hat{h}_{3}\hat{\tilde{N}}_1\hat{\tilde{L}}_3\hat{H}_u+\frac{1}{2}B\hat{M}_{N_1}
\hat{\tilde{N}}_1\hat{\tilde{N}}_1+h.c.)\ea The first line above
corresponds to supersymmetric interactions while the second line
corresponds to soft supersymmetry breaking interactions.

An important consequence of the above lagrangian is that the
lightest sneutrino can be much lighter than the lightest
right-handed neutrino. The reason is that the $2 \times 2$
sneutrino mass-squared matrix is modified because of the
right-handed diagonal mass-squared contribution $(M_R^2 \equiv
\hat{M}_{N_1}^2+\tilde{m}^2)$ and the off-diagonal $A$-term
contribution in (\ref{lightrhneutrino}) (when the higgs gets a
\emph{vev}). RG running from high-scale to low-scale can have
important effects as well. This leads generically to a situation
in which the two mass eigenstates are split by a large amount,
leading to an eigenstate with suppressed mass and couplings
compared to that in the MSSM. The $Z$-width constraint (for
sneutrinos lighter than $m_Z$) can also be satisfied if the
lightest eigenstate is predominantly $\hat{\tilde{N}}_1$. Thus,
the possibility of sneutrino dark matter opens up quite naturally
within gravity mediation. A scenario with similar features for the
right-handed neutrino and sneutrino has been studied in
\cite{ArkaniHamed:2000bq} but from a very different theoretical
perspective. However, the phenomenological consequences are quite
similar for right-handed neutrinos and sneutrinos. From
(\ref{lightrhneutrino}), we see that there is generically also a
$B-L$ violating $B$-term in the soft lagrangian. If for some
reason, this $B$-term is suppressed compared to the $B-L$
conserving mass terms for the sneutrinos, it will split the CP-odd
and CP-even states by a small amount. This could give rise to an
inelastic sneutrino dark matter scenario which could explain the
positive signal at DAMA \cite{Bernabei:2008yi} while being
consistent with the negative results of other direct detection
experiments such as CDMS, XENON, ZEPLIN, KIMS and CRESST. Such a
possibility was pointed out in \cite{TuckerSmith:2004jv}. Of
course, within the theoretical framework considered in this paper,
one has to come up with a natural way of obtaining the correct
splitting. This is left for future work.

Depending on the precise pattern of masses of the right-handed
neutrino and the sneutrinos, there could also be a wide variety of
possibilities for collider physics. For instance, if the
right-handed neutrino is really light - lighter than the SM-like
higgs, then the higgs could preferentially decay to right handed
neutrinos which in turn decay to six particles in the final state
\cite{Graesser:2007pc}. Even if such a decay is kinematically not
allowed, the higgs could decay to $\tilde{\nu}_1\tilde{\nu}_1^*$
since the lightest sneutrino can naturally be quite light as
explained in the previous paragraph. This could give rise to an
invisibly decaying higgs which would make it hard to discover it
at the LHC by standard methods, although other channels may open
up which still make it possible \cite{Frederiksen:1994me}. The
presence of unsuppressed $A$-terms can give non-standard charged
higgs decays, with $H^{\pm}$ decaying to $\tilde{L}_L\tilde{N}_1$.
It could also lead to production of the light ($\tilde{\nu}_1$)
and heavy ($\tilde{\nu}_2$) sneutrinos which could decay either
visibly or invisibly leading to different signatures. In certain
cases, $\tilde{\nu}_2$ could decay to $\tilde{\nu}_1+h$ with a
large branching fraction providing an interesting new way to
produce higgs particles in cascade decays which could be
comparable to standard channels. Thus, having light right-handed
neutrino and sneutrinos opens up a wealth of possibilities at the
LHC. In order to make more concrete predictions however, one needs
to specify an explicit pattern of the other superpartners such as
the gauginos, squarks and the sleptons and then look at the
possible consequences in a systematic manner.

It is important to note that the lightest neutrino mass in this framework is virtually 
massless. This property can also lead to observable signals. For example, the rate for 
neutrinoless double beta decay depends on the effective mass of the electron-type neutrino $m_{\nu_{ee}}$ (for majorana neutrinos). If the light neutrinos follow a hierarchical pattern as in the present framework with the lightest of them being virtually massless, it is possible to bound $m_{\nu_{ee}}$ from both sides. This could lead to a potential observation in the future. This is similar to the arguments in \cite{Fujii:2002hp}.

\section{Conclusions}\label{conclude}

The standard paradigm of cosmological evolution consists of a
radiation dominated era resulting from the decay of the inflaton
which lasts until well after the end of Big-Bang Nucleosynthesis
(BBN). Within this paradigm, DM particles are created from the
thermal plasma and eventually freezeout giving rise to a
\emph{thermal} relic abundance which has to be compared with
observations. Most models trying to explain the baryon asymmetry
also work within the above paradigm. However, as we have argued in
this paper, within classes of realistic string compactifications
which stabilize the moduli as well as generate the Hierarchy at
the same time, the standard paradigm \emph{is no longer
applicable}. This is because in such compactifications, there
generically exists at least one light modulus which is comparable
to the gravitino mass (within a few orders of magnitude). Since the Hubble parameter during inflation $H_{inf}$ is typically much larger than the TeV scale for natural inflationary models, this modulus is generically displaced during inflation. When the hubble parameter drops down
to the mass of the modulus, the modulus starts to oscillate and
quickly dominates the energy density of the Universe. Since the
modulus couples only very weakly to the matter fields, it decays
close to BBN, generating a large entropy and diluting any
pre-existing baryon asymmetry. Thus, most mechanisms explaining
the baryon asymmetry have to take this crucial feature into
account.

In this work, we have argued that it is still possible to generate
the required baryon asymmetry in a natural manner within classes
of realistic string compactifications with a minimal extension of
the MSSM below the unification scale (only right-handed neutrinos)
and satisfying certain microscopic criteria. This is achieved by a
period of Affleck-Dine leptogenesis from the $LH_u$ flat-direction
shortly after the end of inflation, which has been studied earlier
in the literature. In this work, we have embedded the above
mechanism (with some important differences compared to the
original proposals) in a complete framework motivated from string
theory and addressed all relevant issues starting from the end of
inflation to the beginning of BBN.

Since the $LH_u$ flat-direction is obtained after integrating out
heavy right-handed majorana neutrinos; therefore the issue of
neutrino masses is intricately connected to the baryon asymmetry.
In fact, generating the correct baryon asymmetry requires that the
lightest left-handed neutrino mass is virtually massless, of
${\cal O}(10^{-16}$ eV). It is quite interesting that a virtually
massless lightest left-handed neutrino is allowed by data. We have
also argued that it is possible to generate such a light
left-handed neutrino mass naturally from string instantons
(satisfying certain constraints). The moduli and gravitino
problems can be naturally solved in this framework both within
gauge and gravity mediation, although in a different manner.
Because of the decay of heavier moduli before that of the lightest
one, any thermal abundance of DM particles (both in gravity and
gauge mediation) is greatly diluted. Hence, the relic abundance of
DM particles comes from the direct decay of moduli. Since the
decay widths depend on the moduli-matter and moduli-gravitino
couplings, the upper bound on the relic abundance provides an
important constraint on these couplings. Since these couplings are
themselves determined in terms of the structure of the k\"{a}hler
potential among other things, this may provide important insights
about the effective action arising in a particular framework. It
is important to note that it is naturally possible to have
superpartners, particularly gauginos, in the sub-TeV range in this
framework (both within gravity and gauge mediation). The framework also leads
to some broad potential signals for particle physics and cosmology. A
non-thermal origin of Dark Matter, which is quite natural within this framework,
may be crucial to explain the recent results of indirect detection experiments like 
PAMELA and ATIC. Also, a light right-handed (s)neutrino at around
the electroweak scale is naturally allowed (although it is not predicted). This could
have very interesting consequences both for Dark Matter direct detection and the
LHC. Finally, the fact that the lightest neutrino is virtually massless could potentially
lead to a positive signal at neutrinoless double-beta decay experiments.

Each of the microscopic criteria required have separately been
shown to be satisfied for explicit constructions and are mutually
compatible with each other; it is therefore expected that these
criteria could also be satisfied simultaneously by a sizable
fraction within the sub-landscape of realistic string vacua,
providing a solution to all the above problems in a natural
manner. Nevertheless, it would be extremely convincing and useful
if one could have explicit string constructions satisfying most
(if not all) microscopic criteria listed in this paper. This is
technically challenging at present, but could be achieved in the
near future.

It is also worth mentioning that given the existence of light
moduli decaying close to BBN in a string-motivated framework,
there could be other ways of generating the required baryon
asymmetry compared to the one described here. For example, one
could imagine a scenario in which both the baryon asymmetry and
Dark Matter are produced from the decays of a heavy scalar field
dominating the energy density of the Universe before BBN. Such
models have been studied in \cite{Kitano:2008tk}, and it may be
possible to embed them (again maybe with some differences)
naturally in a string theoretic framework \cite{Kumar:2008oc}. In
principle, this mechanism of producing the baryon asymmetry is
distinguishable from the mechanism studied in this paper. For
instance, this may be possible in the presence of light
right-handed (s)neutrinos because of their special signatures for
the LHC and DM. It would be extremely interesting to come up with
other ways of distinguishing the different mechanisms with
experimental observables.

\acknowledgements

PK would like to particularly thank Lawrence Hall for very helpful
discussions. PK is also thankful to Michael Lennek, Gordon Kane
and Scott Watson for useful comments and suggestions. The research
of PK is supported in part by the US Department of Energy.
Finally, PK would like to thank the Aspen Center for Physics and
the TH Unit at CERN for their hospitality where part of the
research was conducted.

\appendix

\section{Lightest ``Modulus" in Gauge Mediation}\label{gaugemodulus}

In this section, we will study the lightest scalar field
(``modulus") in gauge mediation. In accordance with our philosophy
the gauge mediation model should be embedded in a string
compactification, although from a conceptual perspective since
gauge mediation models work at parametrically low energies, the
details of the compactification and string embedding should not
have much effect on low energy physics. Thus, it is reasonable to
expect that a string embedding will stabilize the geometric moduli
with masses much above the TeV scale implying that they will decay
much before BBN even if some of them are displaced from their
minima during inflation. However, there are other scalar fields
(which we also denote by moduli by a slight abuse of notation)
whose effects have to be taken into account. Many models of gauge
mediation exist in the literature, and the precise results for the
masses of light scalar fields will depend on model-dependent
details. However, we would like to argue that there are some
generic features which exist in a large class of gauge mediation
models.

Here we study a simple scheme of gauge mediation at low energies
which was argued
to be quite generic\cite{Ibe:2007km,Murayama:2007fe}: \ba W &=& W_0+\mu^2\,S-\kappa\,S\bar{f}f \nonumber\\
K &=& |f|^2+|\bar{f}|^2+|S|^2-\frac{|S|^4}{\Lambda^2}+... \ea
where $S$ is the goldstino superfield which parameterizes
supersymmetry breaking, $f,\bar{f}$ are the messenger fields
charged under both the visible and hidden gauge groups, $\Lambda$
is a mass scale at which other massive fields have been integrated
out and $W_0$ is the constant piece of the superpotential required
to obtain a vanishing (tiny) cosmological constant. The presence
of $W_0$ breaks $R$-symmetry. A good string embedding should
microscopically account for this constant piece
$W_0$\footnote{after stabilizing the moduli.}. Even though one is
studying a scheme for gauge mediation, one still needs to compute
the potential within supergravity as this is the low energy theory
obtained from string compactifications preserving ${\cal N}=1$
SUSY. Within supergravity, the potential arising from the above
superpotential and the k\"{a}hler potential has a long-lived
metastable supersymmetry breaking minimum at \cite{Kitano:2006wz}:
\ba \langle S \rangle \sim \frac{\Lambda^2}{m_p};\;\langle f
\rangle = \langle \bar{f} \rangle = 0;\nonumber\\ \langle F_S
\rangle \approx \mu^2 \implies m_{3/2} \sim \frac{\mu^2}{m_p}\ea
For $\Lambda \gtrsim 10^{13}$ GeV, this minimum is stable and the
masses of the scalars $S$ and the messengers are given by
\cite{Kitano:2006wz}: \ba m_S \sim
\frac{\mu^2}{\Lambda};\;\;m_{f,\bar{f}} \sim
\frac{\Lambda^2}{m_p}\ea An upper limit for $\Lambda$ is $m_p$. In
order to find out which of the scalars is lighter, one has to
impose the constraint that the mediation of supersymmetry breaking
by gauge interactions dominates compared to that by gravitational
interactions (our original assumption). This leads to an upper
bound on $m_{3/2}$, $m_{3/2} \lesssim {\cal O}(1)$ GeV. Using
$m_p\gtrsim\Lambda \gtrsim 10^{13}$ GeV from above and
conservatively using the upper bound for $m_{3/2}$, one finds that
$m_{f,\bar{f}} > m_S$. Thus, $S$ is generically the lightest
scalar in generic models of gauge mediation with a mass: \ba m_S
\sim \frac{\mu^2}{\Lambda} \sim m_{3/2}\,(\frac{m_p}{\Lambda})\ea
Regarding possible values of $\Lambda$, $\Lambda \sim M_{GUT} \sim
10^{16}$ GeV seems to be both theoretically and phenomenologically
\cite{Ibe:2007km} interesting implying that $m_S \sim
10^2\,m_{3/2}$, although other values in the above range are
presumably allowed. Thus, the lightest scalar in generic gauge
mediation models is a few orders of magnitude above $m_{3/2}$.
Finally, since $S$ is coupled much more strongly to the visible
sector than a gravitationally coupled scalar, it can easily reheat
the Universe to temperatures above a few MeV even for much smaller
masses.

\section{What if $\hat{\tilde{N}}_3$ is displaced during
inflation?}\label{fateofN3}

In this section, we will argue that if $\hat{\tilde{N}}_3$ is
displaced from its true minimum during inflation, then it is not
possible to generate the required baryon asymmetry. During and
after inflation, the dominant (mass-squared) contributions to the
potential for $\hat{\tilde{N}}_3$ are given by: \ba V =
(M_N^2+c_N^2H^2+m_0^2)|\hat{\tilde{N}}_3|^2+[(B+b_NH)\,M_N\,\hat{\tilde{N}}_3\hat{\tilde{N}}_3+h.c.]+....\ea
The parameters $b$ and $B$ can be taken to be real without loss of
generality. The above mass matrix can be diagonalized to give mass
eigenstates: \ba \hat{\tilde{N}}_R &=&
\frac{1}{\sqrt{2}}\,(\hat{\tilde{N}}_3 +
\hat{\tilde{N}}_3^{*})\nonumber\\ \hat{\tilde{N}}_I &=&
\frac{-i}{\sqrt{2}}\,(\hat{\tilde{N}}_3 -\hat{\tilde{N}}_3^{*})\ea
with the following eigenvalues: \ba M^2_{R,I} = (M_N^2+c_NH^2)\pm
(B+bH)\,M_N\ea The situation above is quite different compared to
that with flat-directions because here one has renormalizable $L$
violating interactions (the $B$ term) in contrast to those present
in the case of flat-directions. The $B$ term makes the mass
eigenstates non-degenerate, so they will oscillate independently
with different frequencies when $H \sim M_N$. Therefore, the
lepton number created during these oscillations will oscillate in
general, in contrast to that for flat-directions. Depending on
whether the hubble-induced $B$-term $b$ is ${\cal O}(1)$ or
suppressed (maybe due to a symmetry), one might hope that it is
possible to transfer this lepton number generated for certain
ranges of $\Gamma_N$. This range is determined in terms of of $B$
and $b$ \cite{Allahverdi:2004ix}. However, it turns out that even
though lepton number can be stored in the oscillations of
$\hat{\tilde{N}}_R$ and $\hat{\tilde{N}}_I$ for certain ranges of
$\Gamma_N$, it is still not possible to transfer the lepton
asymmetry to the (s)leptons. This is because the coupling of the
sneutrino to left-handed (s)leptons also violates lepton number
(due to the $B$-term). A simple way to see this is to write the
relevant interaction in the mass eigenstate basis:
\ba\label{interaction} -\mathcal{L}_{int} &=&
\hat{\tilde{N}}_3\,(\hat{h}_1\hat{L}\hat{\bar{\tilde{H}}}_u+\hat{h}_1^*M_N\hat{\tilde{L}}\hat{H_u}^*+
A\,\hat{h}_1\hat{\tilde{L}}\hat{H}_u)+h.c. \nonumber\\ &=&
\frac{\hat{h}_1}{\sqrt{2}}\,\hat{\tilde{N}}_R\left[\hat{L}\hat{\bar{\tilde{H}}}_u+\,(A+M_N)\,\hat{\tilde{L}}\hat{H}_u\right]+h.c
+ \nonumber\\ & &
i\frac{\hat{h}_1}{\sqrt{2}}\,\hat{\tilde{N}}_I\left[\hat{L}\hat{\bar{\tilde{H}}}_u+\,(A-M_N)\,\hat{\tilde{L}}\hat{H}_u\right]+h.c\ea
From (\ref{interaction}), it is clear that the decay widths to
(s)leptons and anti-(s)leptons are the same, and hence no
asymmetry can be generated. Therefore, it is crucial for
$\hat{\tilde{N}}_3$ to not be displaced from its minimum for the
desired baryon asymmetry to be generated.

\section{Technical Details for computing the Lepton
Number}\label{techdetails}

In this section, we will study the potential for the
flat-direction $\phi$ during and after inflation in detail. As
explained in section \ref{modulispectra}, for concreteness we
assume that the visible sector at around the compactification
scale $M_s$ (which we assume to be close to the string scale and
the GUT scale and the same as the $B-L$ breaking scale from
section )\footnote{Although the GUT scale $M_G$ is less than the
typical string scale $M_s$ by a factor of a few, threshold
corrections can account for the discrepancy.} consists of the
gauge group $G=SM \times U(1)$, while the matter sector consists
of that of the MSSM with three right-handed neutrinos and possibly
other (vector-like) exotic fields. These exotic fields typically
get massive at $M_s$. Some of these exotics could be charged under
$U(1)_{B-L}$. This is assumed to be the case. The superpotential,
written schematically in (\ref{W1}), is given in detail by: \ba
\label{W2} \hat{W} =
\hat{h}\hat{N}_3\hat{L}\hat{H}_u+\hat{M}_N\,\hat{N}_3\hat{N}_3+\eta\,X(\psi\bar{\psi}-M_s^2)
\ea Here $\psi,\bar{\psi},X$ are MSSM singlets and
$\psi,\bar{\psi}$ are assumed to be charged (oppositely) under
$U(1)_{B-L}$. The third term in (\ref{W2}) above just provides one
possible mechanism to provide a mass of ${\cal O}(M_s)$ for
$\psi,\bar{\psi},X$, so its precise form is not crucial for us as
long as that is achieved. What is important is the fact that there
are fields charged under $U(1)_{B-L}$ which get a mass when $B-L$
is broken. The full potential is given by: \ba \label{V1}V &=&
V_D+V_{susy}+V_{hubble}+V_{soft} \nonumber\\ &=&
\frac{g_{B-L}^2}{2}\,(|\hat{\tilde{N}}|^2-|\hat{\tilde{L}}|^2+q\psi\partial_{\psi}
\,K-q\bar{\psi}\partial_{\bar{\psi}}
\,K-\frac{1}{4\pi^2}\langle\partial_{T_G}K\rangle)^2+\nonumber\\&
&|\mu\hat{H}_u|^2+|\hat{h}\hat{\tilde{L}}\hat{H}_u|^2+
+|\mu\hat{H}_d+\hat{h}\hat{N}\hat{L}|^2+|\hat{h}\hat{\tilde{N}}\hat{H}_u|^2+|\eta
X\psi|^2+|\eta
X\bar{\psi}|^2+|\eta(\psi\bar{\psi}-M_s^2)|^2+\nonumber\\&&
3H^2(\sum_Y\,(b'_Y)|Y|^2)-H(\sum_{Y}\,c_Y Y W_Y+h.c.)+ V_{soft}\ea
Here $K$ is the K\"{a}hler potential for the moduli and matter
fields (see (\ref{KWf})), $T_m$ is the k\"{a}hler modulus whose
axionic partner shifts under $U(1)_{B-L}$. Since we have assumed
that $T_m$ is stabilized by effects such as higher order
corrections to $K$ or by moduli trapping, the third term in
(\ref{V1}) gives rise to an effective $FI$ parameter,
$\xi_{eff}\equiv -\frac{Q_m}{4\pi^2}\langle\partial_{T_m}K\rangle$
where $Q_m$ is a topological charge. The magnitude of the $FI$
parameter is naturally of ${\cal O}(M_s)$ as will be argued at the
end of this section. The terms in the third line in (\ref{V1})
arise from hubble induced supersymmetry breaking during and after
inflation; $b'_Y$ and $c_Y$ are typically of ${\cal O}(1)$. $Y$
stands for all the relevant fields
$\{X,\psi,\bar{\psi},\hat{\tilde{N}}_3,\hat{\tilde{L}},\hat{H}_u,\hat{H}_d\}$.
The contribution from ordinary hidden sector supersymmetry
breaking $V_{soft}$ is much smaller than the other terms, hence it
is not written explicitly.

The potential (\ref{V1}) is quite complicated. In order to make
progress, it is convenient to integrate out heavy fields. This is
also relevant because field with masses equal to or greater than
the hubble parameter settle down at the bottom of the potential
and track their respective minima during the subsequent evolution.
To this end, we will be interested in the region of field space in
which: \ba\label{restrict} |\mu| &\ll& H \lesssim \hat{M}_N <
|\psi| \lesssim M_s
\nonumber\\
\hat{h}|\phi| &\ll& \hat{M}_N \ea where we have introduced the
$LH_u$ flat-direction $\phi$ defined as:\ba \hat{\tilde{L}}^T=
\frac{1}{\sqrt{2}}\left(\phi\;\,0\right);\;\hat{H}_u^T=\frac{1}{\sqrt{2}}\left(0\;\,\phi\right)\ea
We will see that the solutions obtained are consistent with the
above. In this region of field space, one finds that $\bar{\psi}$,
$X$, $\hat{\tilde{N}}_3$ and $\hat{H}_d$ get heavy masses from the
$F$-term contribution to the potential, i.e. from
$V_{susy}+V_{hubble}+V_{soft}$ in (\ref{V1}). Solving the
equations of motion for these fields, their
vacuum-expectation-values (vevs) are respectively: \ba \langle
\bar{\psi} \rangle &\approx& \frac{M_s^2}{\psi};\,\langle X
\rangle \ll M_s;\, \langle H_d\rangle \ll M_s \nonumber\\
\langle\hat{\tilde{N}}_3\rangle &\approx&
-\frac{\hat{h}\hat{\tilde{L}}\hat{H}_u}{\hat{M}_N}\left(\frac{(1-{\cal
O}(1)\frac{H}{\hat{M}_N})}{(1+\frac{\hat{h}^2(|\hat{{H}}_u|^2+|\hat{\tilde{L}}|^2)}
{|\hat{M}_N|^2}+{\cal
O}(1)\frac{H^2}{|\hat{M}_N|^2})}\right)\nonumber\\ &\approx&
-{\cal O}(1)\frac{\hat{h}\hat{\tilde{L}}\hat{H}_u}{\hat{M}_N}\ea
in accordance with the restrictions on the field-space above. In
the above minimum, $X$ and $\bar{\psi}$ get masses of ${\cal
O}(M_s)$, $\hat{\tilde{N}}_3$ gets a mass of ${\cal O}(\hat{M}_N)$
and $\hat{H}_d$ gets a mass of ${\cal O}(H)$. After integrating
out these fields, and neglecting terms proportional to $\mu$ and
$V_{soft}$\footnote{since these are much smaller than other
terms}, the potential (\ref{V1}), now in terms of $\psi$ and
$\phi$, is given by: \ba\label{V2} V &\approx&
\frac{g_{B-L}^2}{2}\;\left[{\cal
O}(1)|\frac{\hat{h}\phi^2}{\hat{M}_N}|^2-\frac{1}{2}|\phi|^2+q\frac{|M_s|^4}{|\psi|^2}
-q|\psi|^2-\frac{N_F}{4\pi^2}\langle\partial_{T_G}K\rangle\right]^2+
\hat{h}^2|\phi|^4+ \nonumber\\ &&  {\cal
O}(1)\frac{|\hat{h}|^4|\phi|^6}{|\hat{M}_N|^2}+3(b'_{\phi})H^2|\phi|^2+{\cal
O}(1)(3b'_N+1)\,H^2\frac{|\hat{h}|^2|\phi|^4}{|\hat{M}_N|^2}+3(b'_{\psi})H^2|\psi|^2+\nonumber\\
& &3(b'_{\bar{\psi}})H^2\frac{M_s^4}{|\psi|^2}-{\cal
O}(1)H(\frac{\hat{h}^2\phi^4}{\hat{M}_N}+h.c.)\ea The $D$-term
contribution in (\ref{V2}) can vanish naturally by a shift of
$\psi$. Requiring a vanishing $D$-term is justified since the
curvature around the minimum of the $D$-term potential is of
${\cal O}(M_s^2)$, which is much larger than the curvature (${\cal
O}(H^2)$) of the $F$-term potential. This guarantees that $\phi$
remains an approximate flat-direction and gets a large vev, as we
will see below. From (\ref{V2}), the vev of $|\psi|$ is given by:
\ba\label{psi} |\psi|^2 &=&
M_s^2\,[(1-\frac{\beta}{4})^{1/2}-\frac{\beta}{4}]\nonumber\\{\rm
where}\; \beta\,M_s^2 &=& (\frac{1}{2}|\phi|^2-\xi_{eff}-{\cal
O}(1)\,\frac{|\hat{h}|^2|\phi|^4}{|\hat{M}_N|^2})\ea $\beta$ has
to satisfy the constraint that the quantity $(1-\frac{\beta}{4})$
in the square root in (\ref{psi}) is positive, implying that
$\beta < 4$. As will be shown self-consistently, $\beta$ can in
fact be naturally smaller than unity, so that one can expand the
expression for $|\psi|$ in (\ref{psi}) in powers of $\beta/4$.
This implies that after substituting the expression for $|\psi|$
in (\ref{psi}) in (\ref{V1}), one gets the following potential for
$\phi$: \ba V &\approx&  {\cal
O}(1)\frac{|\hat{h}|^4|\phi|^6}{|\hat{M}_N|^2}+3(b'_{\phi})H^2|\phi|^2+{\cal
O}(1)(3b'_N+1)\,H^2\frac{|\hat{h}|^2|\phi|^4}{|\hat{M}_N|^2}-{\cal
O}(1)H(\frac{\hat{h}^2\phi^4}{\hat{M}_N}+h.c.)+\nonumber\\
&
&3H^2M_s^2(b'_{\psi}+b'_{\bar{\psi}})+\frac{9}{8}H^2M_s^2\beta(b'_{\bar{\psi}}-b'_{\psi})
+\frac{3}{128}H^2M_s^2\beta^2(19b'_{\bar{\psi}}-b'_{\psi})\nonumber\\
{\rm or,}\;V&\approx& -
\frac{9}{8}H^2(b'_{\bar{\psi}}-b'_{\psi})\xi_{eff}+\frac{3}{128}\frac{H^2}{M_s^2}(19b'_{\bar{\psi}}-b'_{\psi})\xi^2_{eff}
+\left(3(b'_{\phi})+\frac{9}{16}(b'_{\bar{\psi}}-b'_{\psi})-
\frac{3}{128}\frac{\xi_{eff}}{M_s^2}(19b'_{\bar{\psi}}-b'_{\psi})\right)H^2|\phi|^2\nonumber\\
& &+{\cal
O}(1)H^2|\phi|^4\left(\frac{|\hat{h}|^2}{|\lambda|^2\,M_s^2} [2+
3b'_N+b'_{\psi}-b'_{\bar{\psi}}+\frac{3}{64}\frac{\xi_{eff}}{M_s^2}(19b'_{\bar{\psi}}-b'_{\psi})]
+\frac{3}{512}\frac{(19b'_{\bar{\psi}}-b'_{\psi})}{M_s^2}\right)\nonumber\\
& &-{\cal
O}(1)H(\frac{\hat{h}^2\phi^4}{|\lambda|{M}_s}+h.c.)+{\cal
O}(1)\frac{|\hat{h}|^4|\phi|^6}{|\lambda|^2{M}_s^2}-{\cal
O}(1)\frac{3}{128}(19b'_{\bar{\psi}}-b'_{\psi})\frac{H^2}{M_s^2}\frac{|\hat{h}|^2|\phi|^6}{|\lambda|^2{M}_s^2}\nonumber\\
& &+ {\cal
O}(1)\frac{3}{128}(19b'_{\bar{\psi}}-b'_{\psi})\frac{H^2}{M_s^2}\frac{|\hat{h}|^4|\phi|^8}{|\lambda|^4{M}_s^4}
\ea where we have used $H \lesssim \hat{M}_N$ and $\hat{M}_N \sim
\lambda\,M_s$. From section \ref{result}, one requires a small
$(\frac{\hat{h}^2}{\lambda})$ to get the desired baryon asymmetry.
In addition, we will require a small $(\frac{\hat{h}}{\lambda})$
as well. It has been argued in sections \ref{caseM} and
\ref{caseW} this this can be naturally obtained. This will turn
out to lead to our initial condition $\hat{h}|\phi| \ll
\hat{M}_N$. Therefore, the leading order potential will be
considerably simplified: \ba \label{V3}V &\approx&
V_0+\left(3(b'_{\phi})+\frac{9}{16}(b'_{\bar{\psi}}-b'_{\psi})-
\frac{3}{128}\frac{\xi_{eff}}{M_s^2}(19b'_{\bar{\psi}}-b'_{\psi})\right)H^2|\phi|^2+
{\cal
O}(1)H^2|\phi|^4\left(\frac{3}{512}\frac{(19b'_{\bar{\psi}}-b'_{\psi})}{M_s^2}\right)\ea
where all terms proportional to $(\frac{\hat{h}}{\lambda})$ and
$(\frac{\hat{h}^2}{\lambda})$ arise at subleading order. $V_0$
stands for terms which do not depend on $\phi$. In order to get a
large vev, the mass-squared for $|\phi|$ in (\ref{V3}) has to be
negative. As argued in section \ref{duringinf}, this is quite
naturally possible for $b'_{\phi},b'_{\psi},b'_{\bar{\psi}} \sim
{\cal O}(1)$ and $\xi_{eff} \lesssim M_s^2$. Hence this will be
assumed to be the case. Minimizing the potential with respect to
$|\phi|$, one gets: \ba \label{sol} |\phi|^2 &\approx& {\cal
O}(1)\frac{\left(-3(b'_{\phi})-\frac{9}{16}(b'_{\bar{\psi}}-b'_{\psi})+
\frac{3}{128}\frac{\xi_{eff}}{M_s^2}(19b'_{\bar{\psi}}-b'_{\psi})\right)}
{\left(\frac{3}{256}(19b'_{\bar{\psi}}-b'_{\psi})\right)} \;M_s^2
\equiv \frac{c_{\phi}^2}{2k_{\phi}^2}\,M_s^2 \nonumber\\ &
&\;\;\;\;\;({\rm
using\,the\,notation\,in\,section\,\ref{postinf}}) \ea we see that
$|\phi| \sim M_s$ is naturally allowed. We now check the
self-consistency of our solutions. $\mu \ll H \lesssim \hat{M}_N$
just reflects our starting expectations about low-scale
supersymmetry and high scale inflation together with a large
right-handed neutrino mass $\hat{M}_{N}$. From the solution for
$|\phi|$ above, we have $\frac{\hat{h}|\phi|}{\hat{M}_N} \approx
\frac{\hat{h}}{\lambda}\frac{|\phi|}{M_s} \ll 1$ because
$(\frac{\hat{h}}{\lambda})$ is suppressed. Also, $\xi_{eff}$ is
given by the expression: \ba \xi_{eff} =
-\frac{Q_m}{4\pi^2}\langle
\partial_{T_m}\,K\rangle\ea For $K = m_p^2(-n_m\,\log(T_m+\bar{T}_m)+...)$ with $n_m={\cal O}(1)$ occurring in
string compactifications, one gets: \ba\label{xi} \xi_{eff} \sim
\frac{n_m\,Q_m}{8\pi^2\,\langle\tau_m\rangle}\,m_p^2 \lesssim
M_s^2\ea for $Q={\cal O}(1)$ and $\langle\tau_m\rangle \gtrsim
{\cal O}(1)$. Thus, from (\ref{psi}), (\ref{sol}) and (\ref{xi}),
$\beta < 1$ is quite natural, allowing an expansion of $|\psi|$ in
powers of $\beta/4$. This also implies that $|\psi| \lesssim M_s$
from (\ref{psi}). Thus, we have checked that our solution is
consistent with all requirements on the field space as in
(\ref{restrict}).

\section{$D$-term contribution to masses of Moduli}
\label{dtermmasses}

We saw in section \ref{micro} and in the previous section that
$U(1)$ $D$-terms depend on the moduli. An effective
Fayet-Iliopoulos (FI) parameter arises when these moduli are
stabilized. In order to not destabilize the minima obtained from
the $F$-term potential, one would like all the $D$-terms to
vanish. This will give rise to additional constraints on the
moduli in general. Two kinds of situations can arise. If the
moduli appearing in the $D$-terms are not stabilized by other
effects such as higher order corrections to the k\"{a}hler
potential, then the requirement of a vanishing $D$-term can
stabilize the moduli if the vacuum expectation values (vevs) of
charged matter fields are determined by other considerations. In
the second situation, it could happen that the moduli are
stabilized by other effects such as higher order corrections to
the k\"{a}hler potential. In this case, the {\it vevs} of charged
matter fields could be determined in terms of the stabilized
moduli. For the $U(1)_{B-L}$ $D$-term studied in the previous
section, we have assumed the second case. In both situations, one
could look at fluctuations around the minima of the moduli (where
the $D$-terms vanish) and compute their masses. It turns out that
the $D$-term contribution to the masses of these moduli are
generically much larger than $m_{3/2}$, as we argue below. This is
also supported by arguments given in \cite{Villadoro:2006ia}.

We study the $D$-term contribution to the potential around the
minimum. For concreteness, we will study the $U(1)_{B-L}$ $D$-term
studied in the previous section and compute the mass of the
modulus $T_m$. This is given by: \ba V_D &=& \langle V_D \rangle +
\delta\,V_D\nonumber\\ &=&
\frac{m_p^2\,Q_m^2\,n_m^2}{(8\pi^2)^2\,\langle
\tau_m\rangle^2}\,(\delta\,\hat{T}_m)^2\ea Here we have assumed
that the $D$-term potential vanishes at the minimum and that
$K=m_p^2\,[-n_m\log(T_m+\bar{T}_m)+...]$. Since the mass matrix of
canonically normalized moduli $X_i$ is given by
$\hat{m}^2_{ij}={K}_{ij}^{-1}m_{ij}^2$ where $m_{ij}^2$ is the
mass matrix of the un-normalized moduli, the mass of the
canonically normalized modulus $T_m$ is given by: \ba
\hat{m}_{T_m} &\approx& \frac{n_m^{1/2}\,Q_m\,m_p}{8\pi^2} \gtrsim
10^{-2}\,m_p\ea for $n_m,Q_m={\cal O}(1)$. Thus, the $D$-term
contribution to the moduli masses are much larger than $m_{3/2}$.

\end{document}